\documentclass[12pt]{article}

\usepackage{graphicx,epstopdf,amssymb,amsfonts,amsmath,amsthm,array,
mathrsfs,amscd,flafter,mathtools,cite}

\newcommand{\pbs}[1]{\let\temp=\\#1\let\\=\temp}
\numberwithin{equation}{section}
\def\be{\begin{equation}}\def\ee{\end{equation}}
\def\cvp{\raise 2pt\hbox{,}} 
 \def\tr{\mathop{\rm tr}\nolimits}
  
  \def\im{\mathop{\rm Im}\nolimits}
\def\re{\mathop{\rm Re}\nolimits} \def\diag{\mathop{\rm
diag}\nolimits} 

 \def\d{{\rm d}}\def\nn{{\cal
N}}

 \def\uN{{\text U}(N)}   \def\u{\text{U}(1)} 
\def\b{\text{b}}\def\p{\text{p}}

\def\eff{\text{eff}}
\def\Seff{S_{\text{eff}}}

\def\gs{g_{\text{s}}}
\def\ls{\ell_{\text{s}}}
\def\sd{+}
\def\1{\mathbb{I}}
\def\alp{{\alpha^\prime}}
\def\UK{{\text{U($K$)}}}

\def\AdS{\text{AdS}_{5}}\def\Sfive{\text{S}^{5}}
\def\AdSS{\text{AdS}_{5}\times\text{S}^{5}}

\def\truK{\mathop{\text{tr}_{\text{U}(K)}}\nolimits}

\def\vy{{\vec y\,}}
\def\rr{r_{1}^{2}r_{2}^{2}+ r_{1}^{2}r_{3}^{2}+r_{2}^{2}r_{3}^{2}}

\theoremstyle{plain}

\theoremstyle{definition}
\theoremstyle{remark}

\def\imath#1#2#3{{\it Invent math }{\bf #1} (#2) #3}

\begin{document}
%
%
{\pagestyle{empty}
\parskip 0in

\

\vfill
\begin{center}
  {\LARGE Examples of Emergent Type IIB Backgrounds}

\bigskip

{\LARGE from Matrices}

\vspace{0.4in}

Frank F{\scshape errari}, Micha M{\scshape oskovic}
and Antonin R{\scshape ovai}
\\
\medskip
{\it Service de Physique Th\'eorique et Math\'ematique\\
Universit\'e Libre de Bruxelles and International Solvay Institutes\\
Campus de la Plaine, CP 231, B-1050 Bruxelles, Belgique}



\smallskip

{\tt frank.ferrari@ulb.ac.be, micha.moskovic@ulb.ac.be, antonin.rovai@ulb.ac.be}

\medskip

\end{center}
\vfill\noindent

We study models of emergent space associated with the Coulomb branch, non-commutative and $\beta$ deformations of the $\nn=4$ super Yang-Mills theory, extending a previous work on the undeformed conformal case. The idea is to compute the effective action for D-instantons from the microscopic four-dimensional open-string description and to compare with the non-abelian D-instanton action in the dual ten-dimensional supergravity background. To linear order in the deformation parameter, the D-instantons can probe the full space-time geometry and we can derive all the supergravity fields in this way. We find a perfect match with the known supergravity solutions, including for the Neveu-Schwarz and Ramond-Ramond forms.

\vfill

\medskip
%
\begin{flushleft}
\today
\end{flushleft}
%
\newpage\pagestyle{plain}
\baselineskip 16pt
\setcounter{footnote}{0}

}

\section{Introduction}
\label{IntroSec}

The AdS/CFT correspondence \cite{Maldacena:1997re,Gubser:1998bc,Witten:1998qj} and its generalizations \cite{Aharony:1999ti,D'Hoker:2002aw} offer a framework in which, in principle, geometry and quantum gravity can be studied from well-defined gauge theories without gravity. 
The gauge theories are formulated on the boundary of the bulk space-time. The bulk space emerges together with a metric and other propagating fields from the sum over the gauge theory planar diagrams. This dual ``holographic'' description of the large $N$ limit 
has been used extensively in the literature to understand the strong coupling dynamics of gauge theories. Unfortunately, trying to understand the properties of space and quantum gravity from gauge theory has proven to be much harder \cite{Berenstein:2010xw}. One has to compute at strong coupling in the gauge theory and, even if this were possible, the holographic reconstruction of the geometry from typical field theory correlators is highly non-trivial.

\begin{figure}
\centerline{\includegraphics[width=5in]{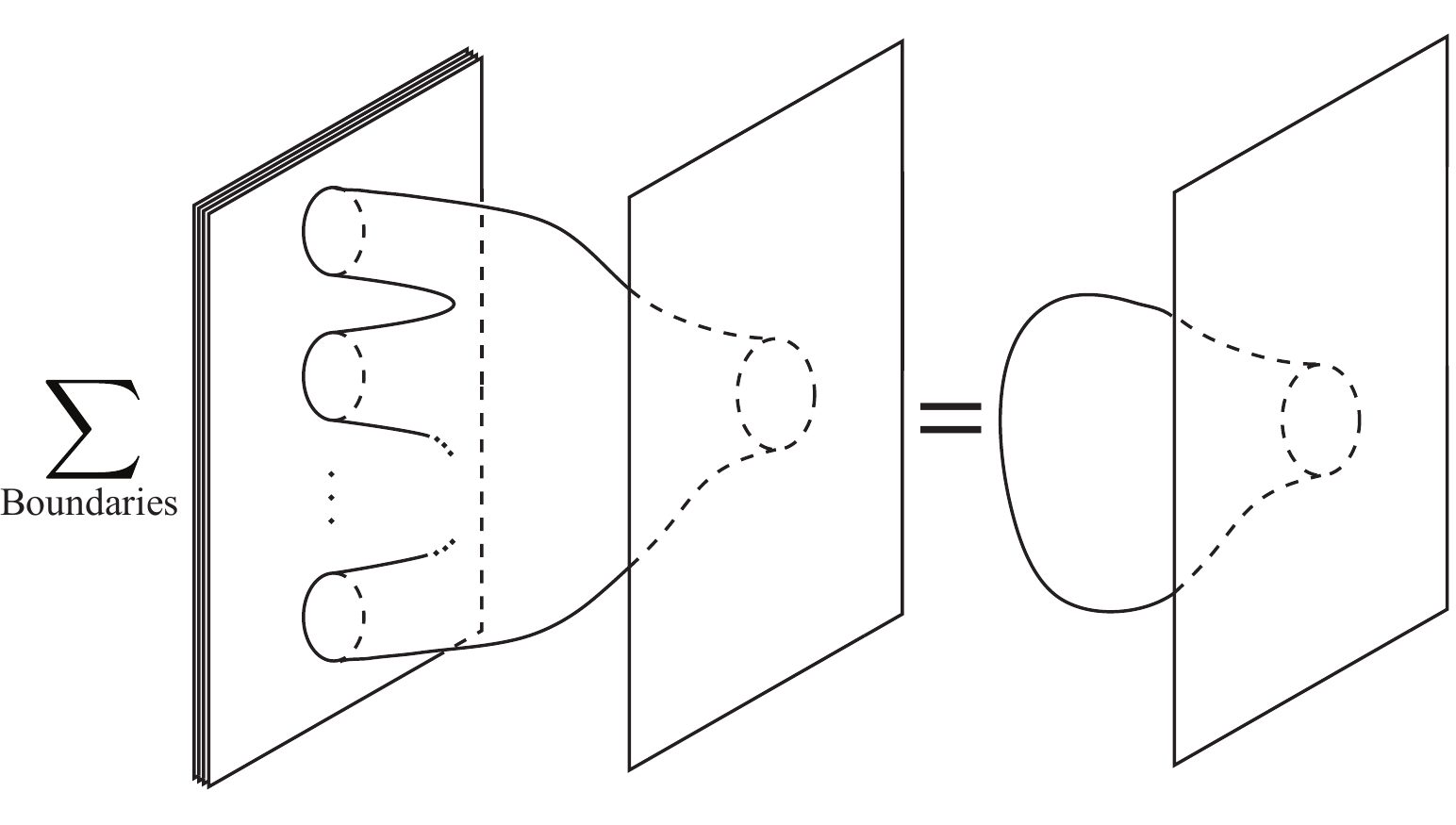}}
\caption{On the left, the world-sheets describing the leading large $N$ interaction between $K$ probe branes and a stack of $N$ background branes. The number of boundaries on the background branes can be arbitrary, corresponding to a sum over loops in the microscopic gauge theoretic path integral. This sum is replaced on the right by a unique open string disk diagram in a non-trivial geometrical background. \label{f1}}
\end{figure}

Recently, one of the authors of the present work proposed a detailed procedure to derive the geometric properties of the bulk geometry from explicitly defined gauge theory correlators \cite{Ferrari:2012nw}. The basic idea is to consider the scattering of $K$ probe branes, $K$ being fixed, off a large number $N$ of background branes, as depicted in Figure \ref{f1}. This system contains three types of open strings, depending on their boundary conditions. The effective action $\Seff$ for the probe branes can be obtained by integrating out the background/background and background/probe open strings. In the usual near-horizon or small $\alpha'$ limit, this amounts to computing a standard gauge-theoretic path integral. Remarkably, the result matches with the non-abelian D-brane action for the probe branes moving in the non-trivial supergravity background created by the background branes. Using known formulas for this non-abelian action \cite{Myers:1999ps}, the supergravity background can then be read off straightforwardly from $\Seff$. 

In \cite{Ferrari:2012nw}, following the above strategy, the full $\AdSS$ background, including the suitably normalized Ramond-Ramond five-form field strength, was derived from a purely field theoretic calculation. The aim of the present paper is to study the emergence of other type IIB geometries from field theory along the same lines, by considering three deformations of the conformal $\nn=4$ super Yang-Mills theory with broken conformal invariance or supersymmetry. 

The simplest deformation we consider is the Coulomb branch deformation, which corresponds to turning on the vacuum expectation values of the scalar fields of the $\nn=4$ theory. This breaks both conformal invariance and R-symmetry but preserves sixteen supersymmetries. The resulting dual geometry asymptotically coincides with the usual $\AdSS$ background in the UV but the metric and the Ramond-Ramond five-form field strength are modified in the IR at the scales set by the scalar expectation values. Our field theory calculations yield a perfect match with the known near-horizon limit of the general multi-centered D3-brane solution, for both the metric and the Ramond-Ramond form.

The second case we consider is the non-commutative deformation \cite{Seiberg:1999vs,Douglas:2001ba}. It breaks conformal invariance but preserves both supersymmetry and R-symmetry. This model does not seem to have a UV fixed point and, accordingly, the known supergravity dual \cite{Hashimoto:1999ut,Maldacena:1999mh} does not have a boundary in the UV and it is likely that a purely field theoretic description does not exist. However, at sufficiently large distance scales, the model approaches the undeformed $\nn=4$ theory and the physical interpretation of both the field theory and its dual supergravity background become clear. The emergent geometry we find is then fully consistent with the background proposed in \cite{Hashimoto:1999ut,Maldacena:1999mh}.

Finally, we investigate the so-called $\beta$-deformation \cite{Leigh:1995ep}. In its most general form \cite{Frolov:2005dj}, it breaks supersymmetry completely but preserves conformal invariance in the planar limit \cite{Ananth:2007px,Ananth:2006ac}. The supergravity solution \cite{Lunin:2005jy,Frolov:2005dj} is known when the deformation parameters are small, which ensures that the $\alpha'$ corrections can be neglected. Again, our solution is fully consistent with supergravity, including for the Neveu-Schwarz and Ramond-Ramond three-form field strengths. Let us note that the form of the dilaton was already derived from instanton calculus in very interesting previous papers (see \cite{Georgiou:2006np,Durnford:2006nb} and related research in \cite{hep-th/9810243,hep-th/9901128,hep-th/9908035,hep-th/9908201,hep-th/9910118,hep-th/0604090,Akhmedov:1998pf}). 

The plan of the paper is as follows. In Section \ref{framesec}, we briefly review the set-up, which is explained in more details in \cite{Ferrari:2012nw}, and present our main results. In particular, we emphasize the new subtleties associated with the use of D-instantons in backgrounds that have a non-constant dilaton \cite{Ferrari:2013pi}. The details of the calculations are included in Section \ref{DerivSec}. We briefly conclude and outline future directions of research in Section \ref{ConclusionSec}. We have also included three appendices containing our conventions, which are consistent with those used in \cite{Ferrari:2012nw}, useful identities and formulas and a review of the supergravity solutions dual to the non-commutative and the $\beta$-deformed models.

\smallskip

\noindent\emph{Note on interesting previous works}

Let us mention other interesting attempts to derive the closed string picture from open string or gauge theory calculations via approaches that complement ours. Beyond the instantons/D-instantons calculations that we have already mentioned \cite{hep-th/9810243,hep-th/9901128,hep-th/9908035,hep-th/9908201,hep-th/9910118,hep-th/0604090,Akhmedov:1998pf,Georgiou:2006np,Durnford:2006nb}, D-instanton corrections to the gravitational background dual to a collection of D-branes have been investigated in \cite{Billo:2011uc,Fucito:2011kb,Billo:2012st}. Similar studies using the boundary state formalism can be found in \cite{Frau:1997mq,DiVecchia:1997pr}. Other fruitful lines of research have been pursued in \cite{Berenstein:2005aa,Berenstein:2005jq,Berenstein:2007wi,Berenstein:2007kq}, where properties of some dual geometries were obtained from the study of matrix models, and in \cite{Steinacker:2010rh,Blaschke:2011qu} in the type IIB matrix model context.

\section{Set-up and main results}
\label{framesec}
\subsection{The general strategy}
\label{strategy}

We consider the path integral for a system of $N\gg1$ background 
D3-branes and $K$ probe D-instantons. In the ``near-horizon,'' 
$\alpha'\rightarrow 0$ limit, this path integral reads
\be\label{pathint1} \int\!\d\mu_{\text{b}}\d\mu_{\text{p}}\,
e^{-S_{\text{b}}-S_{\text{p}}}\, ,\ee
where $S_\b$ is the low energy world-volume action on the D3-branes and $S_\p$ is the action for the D-instanton moduli, taking into account their coupling to the D3-brane local fields. The action $S_\b$ is the $\mathcal N=4$ super Yang-Mills action or a deformation thereof. The action $S_\p$ was derived in \cite{Dorey:2002ik,Green:2000ke,Billo:2002hm}. From the gauge/gravity duality, we expect \eqref{pathint1} to be equivalent to the path integral for $K$ D-instantons in the non-trivial near-horizon closed string background generated by the background branes,
\be\label{pathint2}\int\!\d Z\d\Psi\, e^{-S_{\text{eff}}(Z,\Psi)}\, .\ee
In \eqref{pathint2}, $Z$ and $\Psi$ are the D-instanton matrix bosonic and fermionic moduli in ten dimensions and the effective action $\Seff$ is the non-abelian D-instanton action. This action can be computed \cite{Ferrari:2012nw} from the general formulas given by Myers in \cite{Myers:1999ps}. It depends non-trivially on the supergravity background and thus can be used to obtain the supergravity fields.

In practice, to derive the emerging geometry from $\Seff$, we need to cast
\eqref{pathint1} in the form \eqref{pathint2}. To do this, we first integrate over the D(-1)/D3 string degrees of freedom in \eqref{pathint1}. As explained in \cite{Ferrari:2012nw} (see also \cite{Coleman:1980nk,ZinnJustin:1998cp,Ferrari:2000wq,Ferrari:2001jt,Ferrari:2002gy}), this integration can always be done exactly at large $N$, by introducing auxiliary variables. Some of these variables turn out to correspond to emerging space coordinates,  providing precisely the required moduli to write the result in the form of \eqref{pathint2}. The factor $e^{-\Seff}$ is related to the superdeterminant $\mathscr D$ of a local operator in the four-dimensional gauge theory. Integrating over the D3/D3 strings in \eqref{pathint1} then amounts to computing the expectation value of this superdeterminant,
\be\label{idebasic} \bigl\langle\mathscr D\bigr\rangle (Z,\Psi) = 
e^{-\Seff (Z,\Psi)}\, .
\ee
This equality provides a precise mapping between any state in the D3-brane theory, in which we take the expectation value of $\mathscr D$, and a type IIB geometry, which is encoded in $\Seff$. A crucial point is that the action $\Seff$ obtained in this way will always be proportional to $N$ and thus yields a classical, non-fluctuating emergent geometry at large $N$.

In general, the computation of the expectation value of the superdeterminant $\mathscr D$ involves an intractable sum over planar diagrams.
However, in some interesting cases, drastic simplifications can occur. In particular, for the D(-1)/D3 system under study, it corresponds to a one-point function which cannot be quantum corrected if conformal invariance is unbroken \cite{Ferrari:2012nw}. This is the case, for example, in the planar $\beta$-deformed theory studied below. More generally, we shall assume that when eight or more supercharges are preserved, including when conformal invariance is broken, the expectation value $\langle\mathscr D\rangle$ is not quantum corrected or, more mildly, that the terms in the effective action $\Seff$ that we use to derive the supergravity background are insensitive to the possible quantum corrections in $\langle\mathscr D\rangle$. This is a very plausible assumption, which is strongly supported by the consistency of the results obtained in the present work and in forthcoming publications \cite{emergentD0, emergentD1}. It will be useful to eliminate this caveat in the future and provide a rigorous field theoretic analysis of these non-renormalization properties.

It is important to realize that, even when $\langle\mathscr D\rangle$ is not quantum corrected, the effective action derived from \eqref{idebasic} 
has an explicit non-trivial dependence on the 't~Hooft coupling constant 
\be\label{lambdadef} \lambda = 4\pi\gs N\, ,\ee
coming from the exact integration over the D3/D(-1) degrees of freedom. This integration amounts to summing an infinite class of planar diagrams, with an arbitrary number of loops. We shall see explicit examples in the following. 

When both conformal invariance and supersymmetry are broken, as would be the case, for instance, at finite temperature, the expectation value $\langle\mathscr D\rangle$ will be quantum corrected, yielding an additional and a priori very difficult to compute dependence on $\lambda$. Evaluating the relevance of this contribution will be crucial for future work, see e.g.\ \cite{frankonematrix}.

\subsection{\label{MDactionSec} On Myers' D-instanton action}

To analyse the action $\Seff$, we limit ourselves to the bosonic part, setting $\Psi=0$ in \eqref{idebasic}. We then write the ten $K\times K$ matrices $Z_{M}$, $1\leq M\leq 10$, as
\be\label{Zexp} Z_M=z_M\1+\ls^2 \epsilon_M\ee
and expand $\Seff$ in powers of $\epsilon$,
\begin{equation}
  \Seff=\sum_{n\ge0}\Seff^{(n)}=\sum_{n\ge0}\frac{1}{n!}\ls^{2n}c_{M_1\cdots M_n}(z)\tr\epsilon_{M_1}\cdots\epsilon_{M_n} \, .
  \label{Myersexpansion}
\end{equation}
The coordinates $z_{M}$ correspond to a given ten-dimensional space-time point and we have introduced powers of the string length
\be\label{lsdef} \ls^{2} = 2\pi\alpha'\ee
for convenience. Myers' prescription for the non-abelian D-instanton action yields the coefficients $c_{M_{1}\cdots M_{n}}$ in terms of the supergravity fields, see formula \eqref{Mexpand} in Appendix \ref{appB}. Many terms in \eqref{Mexpand} are actually redundant, being fixed by general consistency conditions \cite{Ferrari:2013pi}. In order to derive the full set of supergravity fields, it is enough to consider the following combinations,
\begin{align}
\label{c0} &c = - 2i\pi\tau = 2i\pi\bigl( C_{0}-ie^{-\phi}\bigr)\\
\label{c3} &c_{[MNP]}  = -\frac{12\pi}{\ls^{2}}\partial_{[M}( \tau B - C_{2})_{NP]}\\
\label{c4} &c_{[MN][PQ]} = -\frac{18\pi}{\ls^{4}}e^{-\phi} 
\bigl(G_{MP}G_{NQ}-G_{MQ}G_{NP}\bigr)\\
\label{c5}
&c_{[MNPQR]}  = -\frac{120 i \pi}{\ls^{4}}
\partial_{[M}\bigl( C_{4} + C_{2}\wedge B - \frac{1}{2}\tau B\wedge B
\bigr)_{NPQR]}\, .
\end{align}

Myers' action has two basic limitations. The first comes from the symmetrized trace prescription \cite{Tseytlin:1997csa,Tseytlin:1999dj} used to fix the ordering ambiguities due to the non-commu\-ting nature of the variables $Z$. This prescription is valid up to order five in the expansion \eqref{Myersexpansion} but is known to fail at higher orders \cite{Bain:1999hu,Sevrin:2001ha}. This caveat will be of no concern to us, since equations \eqref{c0}--\eqref{c5} show that the expansion up to order five is sufficient to fix unambiguously all the supergravity fields. 

The second limitation comes from the fact that the formulas \eqref{c0}--\eqref{c5} are valid only to leading order in the small $\ls^{2}$, or supergravity, approximation. This implies that our microscopic calculations of $\Seff$, which do not rely on a small $\ls^{2}$ approximation, can be compared with Myers' only when $\ls^{2}\rightarrow 0$. When comparing our results with the known supergravity solutions, this restriction is harmless, since the solutions are themselves known at small $\ls^{2}$ only.

Let us point out, however, that some of the basic structural properties of the action, which are visible in the formulas
\eqref{c0}--\eqref{c5}, must be valid to all orders in $\ls^{2}$ because they are consequences of the general consistency conditions discussed in \cite{Ferrari:2013pi}. One of the most interesting properties is that the coefficients $c_{[MNP]}$ and $c_{[MNPQR]}$, viewed as the components of differential forms
\begin{align}\label{F3def} F^{(3)} &= \frac{1}{3!}c_{[MNP]}\,\d z^{M}\wedge\d z^{N}\wedge\d z^{P}\, ,\\ \label{F5def}
F^{(5)} &= \frac{1}{5!}c_{[MNPQR]}\,\d z^{M}\wedge\d z^{N}\wedge\d z^{P}\wedge z^{Q}\wedge z^{R}\, ,\end{align}
must always be closed,
\be\label{F3F5closed} \d F^{(3)} = 0\, ,\quad \d F^{(5)}=0\, .\ee
Locally, we can thus write
\be\label{dc} F^{(3)} = -\frac{4\pi}{\ls^{2}}\,\d C^{(2)}\, ,\quad
F^{(5)} = -\frac{24 i \pi}{\ls^{4}}\,\d C^{(4)}\, .\ee
Since the two- and four-form potentials $C^{(2)}$ and $C^{(4)}$ are well-defined to all order in $\ls^{2}$, formulas \eqref{c3} and \eqref{c5} can actually be used to \emph{define} the Ramond-Ramond and Neveu-Schwarz form fields to all order in $\ls^{2}$, 
\be\label{C2C4NSRR} C^{(2)} = \tau B - C_{2}\, ,\quad C^{(4)} = C_{4} + C_{2}\wedge B -\frac{1}{2}\tau B\wedge B\, ,\ee
modulo the general gauge transformations that are discussed in details in \cite{Ferrari:2013pi}. One of our main goal in the present paper will be to compute the forms \eqref{F3def} and \eqref{F5def} for the Coulomb branch, non-commutative and $\beta$-deformations of the conformal $\nn=4$ gauge theory. As explained in the next Subsection we can then use \eqref{C2C4NSRR} to compare with supergravity in appropriate limits.

Other properties of the Myers action will not, however, be preserved by the $\ls^{2}$ corrections. For example, the only general constraint on the fourth order coefficient $c_{[MN][PQ]}$ is that it should have the same tensorial symmetries as the Riemann tensor. This does not imply a factorization in terms of a second rank symmetric tensor as in \eqref{c4} and thus such a factorization property is generically lost when $\ls^{2}$ corrections are included.

\subsection{\label{useSec}On the use of the non-abelian D-instanton action}

There is one last crucial limitation associated with the use of D-instantons to derive the supergravity background \cite{Ferrari:2013pi}. Intuitively, this limitation is related to the fact that a D-instanton, sitting at a particular point, cannot be expected in general to probe the geometry of the full space-time manifold. This restriction is waived if the effective action, evaluated at $Z_{M}=z_{M}\1$, $\Seff(z\1)=Kc(z)$, does not depend on $z$, or, equivalently,
if the axion-dilaton $\tau$ is constant. This is the case for the $\nn=4$ gauge theory at any point on its Coulomb branch. However, for a generic background with non-constant axion-dilaton, the instantons are forced to sit at the critical points of $c(z)=-2i\pi\tau (z)$. This condition becomes strict when $N\rightarrow\infty$, being equivalent to the saddle-point approximation of the integral \eqref{pathint2}.

An alternative way to understand the same limitation is to study the effect of general matrix coordinate redefinitions on the effective action. It is explained in \cite{Ferrari:2013pi} that, when $\d c$ is generic, one can actually gauge away the coefficients $c_{M_{1}\cdots M_{n}}$ for $n\geq 2$ in the expansion \eqref{Myersexpansion} by an allowed matrix transformation $Z\mapsto Z'$.

For the purposes of the present paper, we shall deal with this difficulty by using a perturbative approach around the $\AdSS$ background on which the instantons can freely move. This is possible because the non-commutative and $\beta$-deformed models are continuous deformations of the $\nn=4$ gauge theory and thus the associated dual backgrounds will be themselves continuous deformations of the $\AdSS$ background. 

Let us denote by $\eta$ the deformation parameter; $\eta$ is the 
dimensionless ratio $\theta/\ls^{2}$ for the non-commutative theory discussed in Section \ref{NCSec} or the combination $\lambda\gamma^{2}$ for the $\beta$-deformed theory studied in Section \ref{BetaSec}. Let us also denote by $\smash{c^{*}_{M_{1}\cdots M_{n}}}$ the coefficients in the expansion \eqref{Myersexpansion} for the undeformed $\AdSS$ background. In our models, the gradient of the axion-dilaton and the corrections to the metric and five-form field strength turn out to be of order $\eta^{2}$. Hence,
\begin{align}\label{adpert} c(z) &= c^{*} + O(\eta^{2})\, ,\\
\label{metpert} c_{[MN][PQ]}(z) &= c_{[MN][PQ]}^{*}(z) + O(\eta^{2})\, ,\\
c_{[MNPQR]}(z) & = c_{[MNPQR]}^{*}(z) + O(\eta^{2})\, , \label{metpert2}
\end{align}
whereas the three-form field strengths are turned on at leading order,
\begin{align}\label{c3pert} c_{[MNP]}(z) = O(\eta)\, .\end{align}
The general variation of $c_{[MNP]}$ under an arbitrary redefinition of the matrix coordinates corresponds to a standard tensorial transformation under diffeomorphisms plus terms proportional to the gradient of $c$ \cite{Ferrari:2013pi} which, by \eqref{adpert}, are $O(\eta^{2})$. \emph{This means that the Neveu-Schwarz and Ramond-Ramond forms $B$ and $C_{2}$ are unambiguously fixed in terms of the microscopic calculation of the coefficient $c_{[MNP]}$ of the D-instanton effective action to leading order in the deformation parameter $\eta$.} 

Moreover, since the background derived from $\Seff$ unambiguously matches with the $\AdSS$ supergravity background \cite{Ferrari:2012nw} in the undeformed theory, we can always choose the same coordinate systems in both points of view at $\eta=0$. In the deformed $\eta\not=0$ models, the coordinate systems $z_{\text{mic}}$ and $z_{\text{SUGRA}}$ used in the effective action $\Seff$ and in the supergravity solution respectively no longer necessarily agree, but the discrepancy must be of order $\eta$,
\be\label{zzequal} z_{\text{mic}}=z_{\textsc{sugra}} + O(\eta)\, .\ee
The associated ambiguity in the axion-dilaton field $c(z)$ is then of order
\be\label{deltac} \delta c = \delta z_{M}\partial_{M} c =O(\eta\partial c) = O(\eta^{3})\, .\ee
\emph{This means that the leading $O(\eta^{2})$ non-constant term in the axion-dilaton field, see \eqref{adpert}, is unambiguously fixed in terms of the microscopic calculation of $c(z)$.}

The conclusion is that, by using D-instantons, we have only access to the leading deformations of the $\AdSS$ background, through the $O(\eta)$ terms in $B$ and $C_{2}$ and the $O(\eta^{2})$ term in $\tau$. Beyond this order, the instantons can no longer probe the full space-time geometry due to the non-trivial dilaton profile. In particular, the backreaction on the metric and five-form cannot be obtained.

Of course, the above restrictions do not apply if we use particles or 
higher-dimensional branes, which can probe the geometry with their kinetic energy. Examples are worked out in \cite{emergentD0,emergentD1}.

\subsection{\label{SetupExamples}The Examples}

We now present our main results, postponing the detailed derivations to the next Section. It is convenient to separate the ten space-time coordinates $(z^{M})$ into four coordinates $(x_{\mu})$ parallel to the background branes and six emergent transverse coordinates $(y_{A})=\vec y$.
The radial coordinate $r$ is defined by
\begin{equation}\label{rdef0}
  r^2=\vy^2\, .
\end{equation}

\subsubsection{The Coulomb branch}

Our first example is the Coulomb branch deformation of the conformal $\text{U}(N)$, $\nn=4$ gauge theory studied in \cite{Ferrari:2012nw}. This deformation is parameterized 
by the scalar expectation values as
\be\label{vevCB}
\left<\varphi_A\right> = \ls^{-2}\, \diag (y_{1A},\ldots,y_{NA})\, ,\quad
1\leq A\leq 6\, .\ee

The supergravity fields, derived from the expansion of the D-instanton effective action computed in Section \ref{CBSec} by comparing with \eqref{c0}--\eqref{c5}, read
\begin{align}\label{taudefCB0}
	\tau &= \frac{4i\pi N}{\lambda}-\frac \vartheta {2\pi}  \cvp \\
\label{metricCB0}	\d s^2 &= H^{-1/2} \d x_\mu \d x_\mu + H^{1/2} \Bigl( \d r^2 + r^2 \d \Omega_5^2 \Bigr) \, , \\\label{F5CB0}
	F_5 &= -\frac{N\ls^4}{\pi R^5}\Bigl( \frac{r^4}{R^4}\, y_{A} \frac{\partial H}{\partial y_{A}}\, \omega_{\Sfive} + i \frac{R^4}{r^4} \, y_{A}\frac{\partial H^{-1}}{\partial y_{A}} \omega_{\AdS} \Bigr) \, .
\end{align} 
We have denoted the metric on the unit round five-sphere by $\d \Omega_5^2$ and used the definitions
\begin{align}\label{defHCB}
  	H(\vec y)& =\frac 1N\sum_{f=1}^N \frac{R^4}{\bigl( \vec y-\vec y_{f}\bigr)^{4}}\, \cvp\\\label{omegaAdSdef}
	\omega_{\AdS}& = \frac{\vy^{2}y_{A}}{R^3}\,
\d x_{1}\wedge\cdots\wedge\d x_{4}\wedge\d y_{A}\, ,\\
\label{omegaS5def}
\omega_{\Sfive} &= \frac{1}{5!}\frac{R^5 y_{F}}{\vy^{6}}\epsilon_{ABCDEF}\,\d y_{A}\wedge\cdots\wedge\d y_{E}\, .
\end{align}
The radius $R$ is related to the string scale and the 't~Hooft coupling $\lambda$ as
\be\label{Rlambdarel} R^{4}=\alpha'^{2}\lambda = \frac{\ls^{4}\lambda}{4\pi^{2}}\, .\ee
The parameter $\vartheta$ is the bare theta angle. The solution \eqref{taudefCB0}, \eqref{metricCB0} and \eqref{F5CB0} matches perfectly the supergravity solution for the multi-centered D3-brane background (a detailed presentation of BPS brane supergravity solutions can be found e.g.\ in \cite{Stelle:1998xg}) in the standard Maldacena scaling limit.

Let us note that the axion-dilaton $\tau$ given by \eqref{taudefCB0} is a constant for the present solution. The D-instantons can thus move freely on the entire space-time geometry and the restriction discussed in \ref{useSec} does not apply. Moreover, the match between the microscopic calculation and the supergravity solution is found at finite $\ls^{2}$ or, equivalently, for 
any value of the 't~Hooft coupling. This suggests that, similarly to the undeformed $\text{AdS}_{5}\times
\text{S}^{5}$ background \cite{Berkovits:2004xu,Kallosh:1998qs,Banks:1998nr}, the near-horizon multi-centered D3-brane background could be exact, with vanishing $\ls^{2}$ corrections to both Myers' action and to the supergravity equations of motion.

Beyond the details of the solution, let us emphasize that general properties like the self-duality of the five-form field strength with respect to the metric \eqref{metricCB0},
\be
	\star F_5 = -i F_5 \, ,
\ee
or the quantization of the five-form flux in units of the D3-brane charge,
\be\label{5formfluxSol1}
	\int _{\vy^{2}=r^{2}} F_5 = 4\pi^2 \ls^4 N(r) \, ,
\ee
where $N(r)$ counts the number of D3-branes with $\vec y_f^{\,2} < r^2$,
which are fundamental consistency requirements from the point of view of the closed string theory, are highly non-trivial and rather mysterious consequences of the microscopic, field theoretic calculation of the effective action. 

\subsubsection{\label{NCsubex}The non-commutative deformation}

Our second example is the non-commutative deformation of the $\nn=4$ gauge theory. This deformation amounts to imposing non-trivial commutation relations among the space-time coordinates \cite{Seiberg:1999vs,Douglas:2001ba}. The most general deformation is parameterized by a real antisymmetric matrix $\theta_{\mu\nu}$, with
\begin{equation}
  [ x_\mu, x_\nu]=-i\theta_{\mu\nu} \, .
  \label{NCrelation}
\end{equation}
Up to an $\text{SO}(4)$ rotation, we may assume that the only non-vanishing components are $\theta_{12}=-\theta_{21}$ and $\theta_{34}=-\theta_{43}$, with corresponding self-dual and anti self-dual parts
\be\label{sdeq} \theta_{12}^{\pm} = \theta_{34}^{\pm} = \frac{1}{2}\bigl(
\theta_{12}\pm\theta_{34}\bigr)\, ,\quad \theta_{\pm}^{2} = \theta^{\pm}_{\mu\nu}\theta^{\pm}_{\mu\nu} = (\theta_{12}\pm\theta_{34})^{2}\, .\ee
As discussed in Section \ref{NCSec}, it can be convenient for some purposes to make the rotation to imaginary Euclidean time $x^{4}\rightarrow ix^{4}$, in which case $\theta_{34}$ is imaginary and $(\theta_{\pm})^{*}=\theta_{\mp}$.

The large $N$ solution of the microscopic model, presented in details in Section \ref{NCSec}, then yields an effective action \eqref{Myersexpansion} with
\be \label{NCc} c=i\vartheta + \frac{8\pi^2 N}\lambda+N
 \left(\sqrt{1+4 \theta_{+}^{2}r^4/R^{8}}-1\right)+N\ln\biggl(\frac{\sqrt{1+4 \theta_{+}^{2}r^4/R^{8}}-1}{2\theta_{+}^{2}r^4/R^{8}}\biggr)\, . 
\ee
Since the coefficient $c$ depends non-trivially on the transverse coordinates $\vec y$, the discussion of Section \ref{useSec} implies that the physical information contained in the effective action is obtained by expanding in $\eta_{\pm}=\theta_{\pm}/\ls^{2}$ around the undeformed $\AdSS$ background. Precisely, \eqref{NCc} can be used to find the axion-dilaton $\tau=i c/(2\pi)$ up to terms of order $\eta^{3}$, giving the predictions
\begin{align}
  C_0&=\frac{\vartheta}{2\pi}-\frac{4i\pi N}{\lambda}\frac{\theta_{12}\theta_{34}}{\ls^4}\frac{r^4}{R^4} +O(\ls^{-2}\theta)^3 \, \cvp \label{C0NCsol}\\
  e^{-\phi}&=\frac{4\pi N}{\lambda}\left[ 1+\left( \left(\frac{\theta_{12}}{\ls^2}\right)^2+\left(\frac{\theta_{34}}{\ls^2}\right)^2 \right)\frac{r^4}{R^4} \right] +O\bigl(\ls^{-2}\theta\bigr)^3 \, .
  \label{phiNCsol}
\end{align}

Moreover, our microscopic calculation yields a third order coefficient $c_{[MNP]}$ and thus a three-form $F^{(3)}$ of the form \eqref{dc}, with a two-form potential $C^{(2)}$ given by
\be
  C^{(2)}=\frac{N\ls^2}{2i\pi\theta_{+}^{2}}\biggl[1- \sqrt{1+4\theta_{+}^{2}r^4/R^{8}}\biggr]\theta^+_{\mu\nu}\,\d x^\mu\wedge\d x^\nu \, .\label{NComega}
\ee
From the discussion of Section \ref{useSec}, we know that only the term linear in the deformation parameter is physical. By using \eqref{C2C4NSRR}, we explain in Section \ref{NCSec} that this yields the prediction
\begin{align}
\begin{split}
  C_2 &=-\frac{r^4}{R^4}\biggl[ \Bigl( \frac{4i\pi N}{\lambda} \frac{\theta_{34}}{\ls^2} + \frac{\vartheta}{2\pi}\frac{\theta_{12}}{\ls^{2}}\Bigr)  
    {\d x_1\wedge\d x_2}\\
    &\hskip 3cm + \Bigl(\frac{4i\pi N}{\lambda}
    \frac{\theta_{12}}{\ls^2} + 
    \frac{\vartheta}{2\pi}\frac{\theta_{34}}{\ls^2}\Bigr)
  {\d x_3\wedge\d x_4} \biggr] +O\bigl(\ls^{-2}\theta\bigr)^2 \, ,
  \end{split}
   \label{C2NCsol} \\
  B& =\frac{r^4}{R^4}\left[ \frac{\theta_{12}}{\ls^2}{\d x_1\wedge\d x_2}+ \frac{\theta_{34}}{\ls^2}{\d x_3\wedge\d x_4}  \right] +
  O\bigl(\ls^{-2}\theta\bigr)^2 \label{BNCsol} \, .
\end{align}

We can now compare the above results with the supergravity solution. This solution was derived independently by Hashimoto and Itzhaki on the one hand \cite{Hashimoto:1999ut} and Maldacena and Russo on the other hand \cite{Maldacena:1999mh}.
As explained previously, to compare the supergravity and microscopic solutions, we must expand in the deformation parameters $\theta_{12}/\ls^{2}$ and $\theta_{34}/\ls^{2}$, which enter into the functions $\Delta_{12}$ and $\Delta_{34}$ defined in \eqref{SolNCDelta}. For the $C_{0}$ field, this expansion plays no r\^ole and indeed equations \eqref{C0NCsol} and \eqref{C0NCsugra} match. For the dilaton field,
we find a match between \eqref{phiNCsol} and \eqref{phiNCsugra} to quadratic order, consistently with our discussion in Section \ref{useSec}.
For the $B$ and $C_{2}$ fields, to compare supergravity with \eqref{BNCsol} and \eqref{C0NCsol}, we must use the approximation $\Delta_{12}\simeq\Delta_{34}\simeq 1$ to keep the leading contribution in the deformation parameter only. We again find a perfect match with the microscopic calculation, in the regime where both can a priori be compared. 

As a final remark, let us note that the dimensionless expansion parameter governing the deformation with respect to the conformal $\nn=4$ model is not really $\eta\sim\theta/\ls^{2}$ but rather the combination 
\be\label{expparamic} \eta_{\text{mic}}=\frac{\theta\, r^2}{R^4}\sim\frac{\theta}{\ls^{2}}\frac{r^{2}}{\ls^{2}\lambda}\ee
in the microscopic formulas \eqref{NCc}, \eqref{NComega} and 
\be\label{expparasugra}\eta_{\textsc{sugra}}= \frac{\theta}{\ls^{2}}\frac{r^{2}}{R^{2}}\sim
\frac{\theta}{\ls^{2}}\frac{r^{2}}{\ls^{2}\sqrt{\lambda}}\ee
in the supergravity solution. In the microscopic formulas, $\lambda$ is a priori arbitrary, but the supergravity solution can be trusted only at large $\lambda$. The condition $\eta_{\textsc{sugra}}\ll 1$ thus automatically implies $\eta_{\text{mic}}\ll 1$ in the supergravity limit. However, 
the condition $\eta_{\textsc{sugra}}\ll 1$ cannot be satisfied for all $r$, 
even if we choose the deformation parameter $\theta/\ls^{2}$ to be arbitrarily small; we have to restrict ourselves to the region $r\ll \ls^{2}\lambda^{1/4}/\theta^{1/2}$, where the solution is indeed a small deformation of the $\AdSS$ background. This means that, even for infinitesimal $\theta$, the theory is completely changed in the UV, a well-known difficulty associated with non-commutative field theories.

\subsubsection{The $\beta$-deformation}

Our last example is the $\beta$-deformed $\nn=4$ gauge theory. The most general deformation studied in Section \ref{BetaSec} is parameterized by three real parameters $\gamma_{1}$, $\gamma_{2}$ and $\gamma_{3}$ and breaks all supersymmetries. Let us discuss here the slightly simpler $\nn=1$ preserving case $\gamma=\gamma_{1}=\gamma_{2}=\gamma_{3}$. In $\nn=1$ language, the $\nn=4$ multiplet decomposes into one vector multiplet and three chiral multiplets $\Phi_{1}$, $\Phi_{2}$ and $\Phi_{3}$. The $\beta$-deformation then simply amounts to replacing the $\nn=4$ preserving superpotential term $\tr [\Phi_{1},\Phi_{2}]\Phi_{3}$ by
$\tr (e^{i\pi\gamma}\Phi_{1}\Phi_{2}\Phi_{3}- e^{-i\pi\gamma}\Phi_{1}\Phi_{3}\Phi_{2})$.

To describe the solution of the model it is convenient to introduce the polar coordinates $(\rho_{i},\theta_{i})$, $1\leq i\leq 3$, defined in terms of the transverse coordinates $\vec y$ by
\begin{alignat}{3}\nonumber y_{1}&= \rho_{1}\cos\theta_{1}\, ,\quad &
y_{3} &=\rho_{2}\cos\theta_{2}\, , \quad & y_{5} &=\rho_{3}\cos\theta_{3}\, ,
\\\label{polardef0}
y_{2} &=\rho_{1}\sin\theta_{1}\, ,&
y_{4} &=\rho_{2}\sin\theta_{2}\, ,& y_{6} &=\rho_{3}\sin\theta_{3}\, ,
\end{alignat}
together with
\be\label{ridef0} r_{i} =\frac{\rho_{i}}{\sqrt{\rho_{1}^{2}+\rho_{2}^{2}+\rho_{3}^{2}}}=\frac{\rho_{i}}{|\vec y|}\,\cvp\ee
which satisfy the constraint 
\be\label{ricons} r_{1}^{2}+r_{2}^{2}+r_{3}^{2}=1\, .\ee
We shall also use the spherical angles $(\theta,\phi)$ defined by
\be\label{thetaphidef0} r_{1}=\sin\theta\cos\phi\, ,\quad r_{2}=\sin\theta\sin\phi\, ,\quad r_{3}=\cos\theta\, .\ee
The large $N$ solution of the microscopic theory, derived in Section \ref{BetaSec}, yields
\be\label{Betac} c = \frac{8\pi^{2}N}{\lambda} + i\vartheta - N \ln\bigl(
1-4(\rr)\sin^{2}(\pi \gamma)\bigr)\, .\ee
Expanding to second order in the deformation parameter $\gamma$ as required by the discussion in Section \ref{useSec}, we obtain the prediction
\be\label{phiBetasol} e^{-\phi} = \frac{4\pi N}{\lambda} 
\Bigl( 1+\frac{1}{2}\lambda\gamma^{2}\bigl(\rr\bigr) + O\bigl(\lambda\gamma^{4}\bigr)
\Bigr)\, .\ee
Moreover, the two-form $C^{(2)}$ defined in \eqref{dc} is found to be
\begin{multline}\label{omegaBetasol}
C^{(2)} = \frac{4 N\ls^{2}}{\pi}\sin\bigl(2\pi\gamma\bigr)\biggl[ G_{1}\wedge\d\theta_{1}
+ G_{2}\wedge\d\theta_{2}+ G_{3}\wedge\d\theta_{3}\\
-\frac{i}{4}\frac{r_{1}^{2}r_{2}^{2}\d\theta_{1}\wedge\d\theta_{2}
+ r_{1}^{2}r_{3}^{2}\d\theta_{1}\wedge\d\theta_{3}
+r_{2}^{2}r_{3}^{2}\d\theta_{2}\wedge\d\theta_{3}}{1-4(\rr)\sin^{2}(\pi\gamma)}\biggr]\, ,
\end{multline}
with
\begin{align}\label{G1def0} \d G_{1} &= \frac{r_{1}r_{2}r_{3}\bigl(r_{1}^{2}
+ (r_{2}^{2}+r_{3}^{2})\cos (2\pi\gamma)\bigr)}{\bigl(1-4(\rr)\sin^{2}(\pi\gamma)\bigr)^{2}}\,\sin\theta\, \d\theta\wedge\d\phi\, ,\\
\label{G2def0} \d G_{2} &= \frac{r_{1}r_{2}r_{3}\bigl(r_{2}^{2}
+ (r_{1}^{2}+r_{3}^{2})\cos (2\pi\gamma)\bigr)}{\bigl(1-4(\rr)\sin^{2}(\pi\gamma)\bigr)^{2}}\,\sin\theta\, \d\theta\wedge\d\phi\, ,\\
\label{G3def0} \d G_{3}& = \frac{r_{1}r_{2}r_{3}\bigl(r_{3}^{2}
+ (r_{1}^{2}+r_{2}^{2})\cos (2\pi\gamma)\bigr)}{\bigl(1-4(\rr)\sin^{2}(\pi\gamma)\bigr)^{2}}\,\sin\theta\, \d\theta\wedge\d\phi\, .
\end{align}
To obtain a prediction for $B$ and $C_{2}$, we are instructed by the discussion in Section \ref{useSec} to expand to linear order in the deformation parameter $\gamma$. In this limit,
\be\label{approxGs}
\d G_{1}\simeq \d G_{2}\simeq\d G_{3} \simeq r_{1}r_{2}r_{3}\,
\sin\theta\,\d\theta\wedge\d\phi = \d\omega_{1}\ee
and \eqref{C2C4NSRR} then yields
\begin{align} \label{C2Betasol} C_{2}&=-8 N\ls^{2}\gamma\omega_{1}\wedge
\bigl(\d\theta_{1}+\d\theta_{2}+\d\theta_{3}\bigr) + O\bigl(\gamma^{2}\bigr)\, ,
\\
\label{BBetasol} B & =-\frac{\ls^{2}\lambda}{2\pi}\,\gamma\bigl(
r_{1}^{2}r_{2}^{2}\d\theta_{1}\wedge\d\theta_{2}
+ r_{1}^{2}r_{3}^{2}\d\theta_{1}\wedge\d\theta_{3}
+r_{2}^{2}r_{3}^{2}\d\theta_{2}\wedge\d\theta_{3}\bigr) + O\bigl(\gamma^{2}\bigr)\, .
\end{align}

The supergravity dual of the $\beta$-deformed theory was studied by Lunin and Maldacena in \cite{Lunin:2005jy} (or, more generally when $\gamma_{1}$, $\gamma_{2}$ and $\gamma_{3}$ are distinct, by Frolov in \cite{Frolov:2005dj}, see Section \ref{BetaSec}). This is reviewed in Appendix \ref{CCC}. The supergravity solution can be trusted as long as the two conditions
\be\label{sugracondBeta} \lambda\gg 1\, ,\quad \lambda\gamma^{4}\ll 1\, ,\ee
are satisfied. The discussion in Section \ref{useSec} implies that supergravity can be compared with the above microscopic solution only when the background is a small perturbation of the undeformed $\AdSS$ solution. This occurs when $\lambda\gamma^{2}\ll 1$, in which case the functions $1/\sqrt{G}$ and $\sqrt{G}$ in equations \eqref{tbeta} and \eqref{Bbeta} can be simplified. This yields a perfect match with \eqref{phiBetasol}, \eqref{C2Betasol} and \eqref{BBetasol}.

\section{Derivation of the solutions\label{DerivSec}}

Our starting point is the microscopic probe action $S_{\text p}$ for $K$ D-instantons in the undeformed conformal $\nn=4$ model. This action was presented in details in \cite{Ferrari:2012nw} and corresponds to the standard sigma model for the ADHM instanton moduli. Using notations explained in the Appendix \ref{NotAppSec}, it reads
\begin{multline}\label{Sp1} S_{\text p}=K\Bigl(\frac{8\pi^{2}N}{\lambda}
 +i\vartheta\Bigr)+ \frac{4\pi^{2}N}{\lambda}\truK \Bigl\{
2iD_{\mu\nu}\bigl[X_{\mu},X_{\nu}\bigr] -
\bigl[X_{\mu},\phi_{A}\bigr]\bigl[X_{\mu},\phi_{A}\bigr]\\
-2  \Lambda^{\alpha}_{\ a}\sigma_{\mu\alpha\dot\alpha}
\bigl[X_{\mu},\bar\psi^{\dot\alpha a} \bigr]
- \bar\psi_{\dot\alpha}^{\ a}\Sigma_{Aab}
\bigl[\phi_{A},\bar\psi^{\dot\alpha b}\bigr]\Bigr\}\\\hskip 3cm
+ \frac{i}{2}
\tilde q^{\alpha}D_{\mu\nu}\sigma_{\mu\nu\alpha}^{\hphantom{\mu\nu\alpha}\beta}q_{\beta}+\frac{1}{2}\tilde q^{\alpha}\phi_{A}\phi_{A}q_{\alpha}-\frac{1}{2}
\tilde\chi^{a}\Sigma_{Aab}\phi_{A}\chi^{b}\\+ \frac{1}{\sqrt{2}}
\tilde q^{\alpha}\Lambda_{\alpha a}\chi^{a} + \frac{1}{\sqrt{2}}
\tilde\chi^{a}\Lambda^{\alpha}_{\ a}q_{\alpha} + \cdots \end{multline}
The $\cdots$ represent couplings with the local fields of the $\nn=4$ gauge theory living on the background D3-brane worldvolume. These terms are described in \cite{Ferrari:2012nw,Green:2000ke,Billo:2002hm} and enter crucially into the computation of the expectation value \eqref{idebasic} of the superdeterminant $\mathscr D$, but play no r\^ole whatsoever when this determinant is not quantum corrected. As discussed in Section \ref{strategy}, we can thus discard them for our present purposes. The moduli in \eqref{Sp1} organize themselves into a vector multiplet $(\phi_{A},\Lambda_{\alpha a},D_{\mu\nu})$ of six-dimensional $\nn=1$ supersymmetry and an adjoint $(X_{\mu},\bar\psi^{\dot\alpha a}$) and fundamentals $(q_{\alpha},\chi^{a},\tilde q^{\alpha},\tilde\chi^{a})$ hypermultiplets. Their detailed symmetry properties are presented in the Appendix, Table \ref{indices}. Note in particular that the modulus $D_{\mu\nu}$ is self-dual, $D_{\mu\nu}=D_{\mu\nu}^{+}$.

The fields in the vector multiplet $(\phi_{A},\Lambda_{\alpha a},D_{\mu\nu})$ are auxiliary fields that can be easily integrated out from \eqref{Sp1} to yield the usual ADHM constraints and measure on the instanton moduli space. However, keeping these variables is crucial to solve the model at large $N$. In particular, the action, as written in \eqref{Sp1}, is quadratic in the hypermultiplet fields, a property that would be lost if we integrate out the six scalars $\phi_{A}$. Instead, we can integrate exactly over the moduli $q,\tilde q,\chi, \tilde\chi$ which belong to the fundamental of $\uN$. This yields an effective action which is automatically proportional to $N$ and can thus be treated classically when $N\rightarrow\infty$.

The microscopic actions for the deformed theories that we study in the present paper are simple modifications of \eqref{Sp1} and their large $N$ limit can be studied along the same lines. Since our goal is to obtain the bosonic effective action, we shall always set $\Lambda_{\alpha a}$ and $\bar\psi^{\dot\alpha a}$ to zero in the following. We also introduce the notation
\be\label{defY} Y_{A}=\ls^{2}\phi_{A}\, ,\ee
since the auxiliary fields $\phi_{A}$ will turn out, as in \cite{Ferrari:2012nw,hep-th/9810243,hep-th/9901128}, to play the r\^ole of the six emerging transverse coordinates.

\subsection{The Coulomb branch deformation}
\label{CBSec}
\subsubsection{The microscopic action}

The Coulomb branch deformation amounts to turning on non-zero expectation value $\langle\varphi_{A}\rangle$ for the $\nn=4$ scalars. 
The microscopic action is then modified by making the replacement
\be\label{CBmod} \phi_{Ai}^{\ \ j}\delta_{f}^{f'}\rightarrow
\phi_{Ai}^{\ \ j}\delta_{f}^{f'} -\langle\varphi_{Af}^{\ \ f'}\rangle
\delta_{i}^{j} = \phi_{Ai}^{\ \ j}\delta_{f}^{f'} - \ls^{-2}y_{fA}\delta_{f}^{f'}\delta_{i}^{j}\ee
in the third line of \eqref{Sp1}. We have indicated all the $\uN$ and $\text{U}(K)$ indices explicitly for clarity. This modification is actually best understood as coming from the coupling of the scalar fields 
$\varphi_{A}$ to the moduli in the $\cdots$ part of the action \eqref{Sp1} that we have not written down explicitly.

\subsubsection{The effective action}

Integrating out $q,\tilde q,\chi,\tilde\chi$ yields the effective action
\begin{multline}\label{SeffboseCB} S_{\text{eff}}(X,Y,D)=K\Bigl(\frac{8\pi^{2}N}{\lambda}
 +i\vartheta\Bigr)\\+
\frac{4\pi^{2}N}{\ls^{4}\lambda}\truK \Bigl\{
2i\ls^{4} D_{\mu\nu}\bigl[X_{\mu},X_{\nu}\bigr] -
\bigl[X_{\mu},Y_{A}\bigr]\bigl[X_{\mu},Y_{A}\bigr]\Bigr\}+ \ln\Delta_{q,\tilde q} - \ln\Delta_{\chi,\tilde\chi}\, .
\end{multline}
The logarithm of the superdeterminant $\ln(\Delta_{q,\tilde q}/\Delta_{\chi,\tilde\chi})$ is the sum of the term obtained by integrating over the bosonic variables $q,\tilde q$,
\be\label{Deltaqq} \ln\Delta_{q,\tilde q} = \sum_{f=1}^N \ln \det \Bigl(\big(Y_A-y_{f A}\big)^2\otimes \1_{2\times2} + i \ls ^4 D_{\mu\nu}\otimes \sigma_{\mu\nu} \Bigl) \ee
and the term obtained by integrating over the fermionic variables $\chi,\tilde\chi$,
\be\label{Deltacc} -\ln\Delta_{\chi,\tilde\chi} =-\sum_{f=1}^N
\ln \det \bigl( \Sigma_A \otimes \big( Y_A - y_{f A} \big) \bigl)\, .\ee
This action is proportional to $N$ and thus can be treated classically at large $N$. In particular, the fluctuations of
$X$, $Y$ and $D$ are suppressed. The six matrices $Y_{A}$ are interpreted as the six coordinates for the emerging space transverse to the background D3-branes. Together with the four $X_{\mu}$s, 
they correspond to the ten matrix coordinates $Z_{M}$ in the non-abelian D-instanton action \eqref{Myersexpansion}. Consequently, to compare \eqref{SeffboseCB} with \eqref{Myersexpansion}, we simply need to integrate out the additional variables $D_{\mu\nu}$ by solving the saddle-point equation
\be\label{saddleCBD} \frac{\partial\Seff}{\partial D_{\mu\nu i}^{\ \ \ j}}
= 0\ee
and plugging the solution $D_{\mu\nu}=\langle D_{\mu\nu}\rangle$ back into \eqref{SeffboseCB},
\be\label{SeffNClast} \Seff (X,Y) = \Seff \bigl(X,Y,\langle D\rangle\bigr)\, .\ee

Our goal is to expand $\Seff(X,Y)$ as in \eqref{Myersexpansion}, up to the fifth order and then use \eqref{c0}--\eqref{c5} to read off the supergravity background. This calculation is very similar to the one performed in \cite{Ferrari:2012nw}. We set
\be\label{XYexp}
X_\mu = x_\mu \1 + \ls^2 \epsilon_\mu \, , \quad Y_A = y_A \1 + \ls^2 \epsilon_A
\ee
and solve \eqref{saddleCBD} perturbatively in $\epsilon$. Using the standard notation $[\epsilon_\mu,\epsilon_\nu]^+$ for the self-dual part of the commutator (see \eqref{sddef}) and defining the function
\be\label{defH}
	H(\vec y) =\frac 1N\sum_{f=1}^N \frac{R^4}{\bigl( \vec y-\vec y_{f}\bigr)^{4}}\, \cvp
\ee
where $R$ is given by \eqref{Rlambdarel},
we obtain
\be\label{Dorder3CB}
\langle D_{\mu\nu} \rangle= iH^{-1} \left[ \epsilon_\mu,\epsilon_\nu \right]^+ + \frac {i\ls^2}{2}\partial_A H^{-1}\left( \epsilon_{A} [\epsilon_\mu,\epsilon_\nu]^+ + [\epsilon_\mu,\epsilon_\nu]^+ \, \epsilon_{A}\right) +  O\bigl(\epsilon^{4}\bigr) \, .
\ee
Let us note that since $\langle D \rangle$ solves the equation of motion \eqref{saddleCBD}, it enters into \eqref{SeffboseCB} at order $\langle D\rangle^{2}$ and thus the expansion \eqref{Dorder3CB} to third order in $\epsilon$ is sufficient to get the expansion of \eqref{SeffboseCB} to fifth order.

Plugging \eqref{Dorder3CB} into $\eqref{SeffboseCB}$, expanding the determinants by using the relation
\be\label{detexp} \ln\det (M+\delta M) = \ln\det M + \sum_{n\geq 1}
\frac{(-1)^{n+1}}{n}\tr (M^{-1}\delta M)^{n}\ee
and computing the resulting traces by using the identities \eqref{2sigmaid} and \eqref{s1}--\eqref{s5} in the Appendix, we find that the first, second and third order action in \eqref{Myersexpansion} vanish, due to many cancellations between the bosonic and fermionic contributions \eqref{Deltaqq} and \eqref{Deltacc},
\be\label{Sefforder123CB}S_\eff ^{(1)}=S_\eff ^{(2)}=S_\eff ^{(3)}=0\, .\ee
On the other hand, the action is non-trivial at the fourth and fifth orders, %
\begin{align}\label{Sefforder4CB}\begin{split}
\Seff^{(4)}&=-\frac{\ls^8}{2R^4}\tr_\UK\Bigl\{2[\epsilon_A,\epsilon_\mu][\epsilon_A,\epsilon_\mu]+H^{-1}[\epsilon_\mu,\epsilon_\nu][\epsilon_\mu,\epsilon_\nu] \\& \hskip 8cm+ \frac 12 H [\epsilon_A,\epsilon_B][\epsilon_A,\epsilon_B]\Bigr\}\, ,\end{split} \\
\label{Sefforder5CB}
\begin{split}
\Seff^{(5)} &=-\frac{\ls^{10}}{2R^4}\partial_A H^{-1} \tr_\UK \Bigl\{ \epsilon_A [\epsilon_\mu,\epsilon_\nu][\epsilon_\mu,\epsilon_\nu] + 2 \epsilon_{\mu\nu\rho\lambda} \epsilon_A \epsilon_\mu \epsilon_\nu \epsilon_\rho \epsilon_\lambda\\ &\hskip 2.5cm
-H^2 \epsilon_A [\epsilon_B,\epsilon_C][\epsilon_B,\epsilon_C] 
-\frac{2iH^2}{5}\epsilon_{ABCDEF}\epsilon_{B}\epsilon_{C}\epsilon_{D}\epsilon_{E}\epsilon_{F} \Bigr\}\,.
\end{split}
\end{align}

\subsubsection{The emergent geometry}

The results of the previous subsection are perfectly consistent with the general ideas explained in Section \ref{framesec} and depicted in Figure \ref{f1}. The effective action that we have obtained can be matched with the non-abelian action for D-instantons embedded in a non-trivial ten-dimensional emergent geometry, with background supergravity fields fixed by comparing 
\eqref{Sefforder123CB}, \eqref{Sefforder4CB} and \eqref{Sefforder5CB} with \eqref{Mexpand} or equivalently \eqref{c0}--\eqref{c5}.

The conditions $\Seff^{(1)}=\Seff^{(2)}=0$ imply that the axion-dilaton is a constant,
\be\label{tauCB}
	\tau =  \frac{4i\pi N}{\lambda}-\frac \vartheta {2\pi}\, \cvp
\ee
whereas $\Seff^{(3)}=0$ yields
\be\label{3formCB} B=C_{2}=0\, .\ee
On the other hand, the fourth order term \eqref{Sefforder4CB} allows to identify the coefficient $c_{[MN][PQ]}$ which turns out to be precisely of the required form \eqref{c4}, with a metric
\be\label{MetricCB}
G_{\mu\nu}=H^{-1/2} \delta_{\mu\nu} \, , \quad G_{AB}=H^{1/2}\delta_{AB} \,, \quad G_{A\mu}=0
\ee
which is equivalent to \eqref{metricCB0}. Finally, we get the completely antisymmetric coef\-fi\-cient $c_{[MNPQR]}$ from \eqref{Sefforder5CB}, which yields the five-form field strength by comparing with \eqref{c5} and using \eqref{3formCB},
\be\label{5FormCB}
({F_5})_{ABCDE} = -\frac{N\ls^4}{\pi R^4} \partial_F H \epsilon_{ABCDEF} \,,\quad
({F_5})_{A\mu_1\cdots \mu_4}=-\frac{iN\ls^4}{\pi R^4} \partial_A H^{-1} \epsilon_{\mu_1 \cdots \mu_4}\, ,
\ee
and all the other independent components (not related to \eqref{5FormCB} by antisymmetry) vanishing. This is equivalent to the formula \eqref{F5CB0}.

\subsection{The non-commutative deformation}
\label{NCSec}
\subsubsection{The microscopic action}

The non-commutative deformation can be elegantly implemented by replacing all ordinary products $fg$ appearing in the microscopic action by the so-called Moyal $*$-product defined by
\be\label{Moyaldef} f*g = e^{-\frac{i}{2}\theta_{\mu\nu}P^{f}_{\mu}P^{g}_{\nu}}\cdot (fg)\, ,\ee
where $P_{\mu}^{f}$ and $P_{\mu}^{g}$ are the translation operators acting on $f$ and $g$ respectively and $\theta_{\mu\nu}$ is an arbitrary antisymmetric matrix \cite{Seiberg:1999vs,Douglas:2001ba}. The only moduli in \eqref{Sp1} transforming non-trivially under translations are the matrices $X_{\mu}$, with $P_{\mu}\cdot X_{\nu} = -i\delta_{\mu\nu}$. It is then easy to check that the only term affected by the use of the $*$-product is the commutator term
\be\label{stareffectNC}\tr D_{\mu\nu} [X_{\mu},X_{\nu}] \rightarrow\tr D_{\mu\nu}\bigl( X_{\mu}*X_{\nu}-
X_{\nu}*X_{\mu}\bigr) =\tr D_{\mu\nu}\bigl( [X_{\mu},X_{\nu}] + i\theta_{\mu\nu}\bigr)\, .\ee
This simple reasoning reproduces the well-known modification of the ADHM construction in non-commutative gauge theories \cite{Nekrasov:1998ss}. Note that, in particular, the action only depends on the self-dual part 
$\theta_{\mu\nu}^{+}$ of the non-commutative parameters because the modulus $D_{\mu\nu}$ is itself self-dual.

\subsubsection{The effective action\label{effactNC}}

Integrating out $q,\tilde q,\chi,\tilde\chi$ from the microscopic action yields
\begin{multline}\label{SeffboseNC} S_{\text{eff}}(X,Y,D)=K\Bigl(\frac{8\pi^{2}N}{\lambda}
 +i\vartheta\Bigr) -\frac{4\pi^{2}N}{\ls^{4}\lambda}\tr
 \bigl[X_{\mu},Y_{A}\bigr]\bigl[X_{\mu},Y_{A}\bigr] \\ - 
N \ln \det \bigl( \Sigma_A \otimes Y_A \bigl)
 + \mathcal S\bigl([X_\mu,X_\nu]^+,\vec Y^2, D;\theta_+\bigr)\, .
\end{multline}
We have singled out the $D$-dependent piece in the action,
\begin{multline}\label{calSNC}
\mathcal S\bigl([X_\mu,X_\nu]^+,\vec Y^2,D\bigr)=\frac{8i\pi^{2}N}{\lambda}\tr D_{\mu\nu}\bigl(\bigl[X_{\mu},X_{\nu}\bigr]+i\theta_{\mu\nu}\bigr)^+ \\+ N\ln \det \bigl(\vec 
Y^{2}\otimes \1_{2} + i \ls ^4 D_{\mu\nu}\otimes \sigma_{\mu\nu} \bigl)\, .
\end{multline}
Let us note that the determinants appearing in \eqref{SeffboseNC} and \eqref{calSNC} are special cases of the determinants \eqref{Deltaqq} and \eqref{Deltacc} studied in the previous subsection. The crucial difference comes from the saddle-point equation \eqref{saddleCBD}, which now picks 
a new term in $\theta_{\mu\nu}$,
\begin{equation}
  \frac{\partial \mathcal S}{\partial D_{\mu\nu j}^{\ \ \ i}}=\frac{8\pi^2}\lambda\Bigl( [X_\mu,X_\nu]^{+\,j}_{\ i}+i\theta_{\mu\nu}^{+}
  \delta_{i}^{j} \Bigr)+\ls^4 
  \Bigl(\vec Y^{2}\otimes\1_{2}+i\ls^4D_{\rho\kappa}\otimes\sigma_{\rho\kappa}\Bigr)^{-1\ j\beta }_{\ i\alpha }\sigma_{\mu\nu\beta}^{\ \ 
\ \alpha}=0\, .
  \label{EOMDNC}
\end{equation}
This equation must be solved for $D_{\mu\nu}=\langle D_{\mu\nu}\rangle$, order by order in the expansion \eqref{XYexp}. 

By using \eqref{inv1sigma}, we find a quadratic equation for the zeroth order solution. Picking the root that behaves smoothly when $\theta_{\mu\nu}\rightarrow 0$ yields
\be\label{solD0NC} \langle D_{\mu\nu}\rangle = \frac{\lambda}{8\pi^{2}\theta_{+}^{2}}\biggl( 1 - \sqrt{1+4\theta_{+}^{2}r^{4}/R^{8}}
\biggr) \theta_{\mu\nu}^{+} + O\bigl(\epsilon\bigr)\ee
in terms of the transverse radial coordinate \eqref{rdef0} and the parameter $\theta_{+}$ defined in \eqref{sdeq}. Plugging this result into \eqref{calSNC} and \eqref{SeffboseNC} and computing the determinants using \eqref{detsigmaid} and \eqref{detsigmaa}, we get the zeroth order coefficient \eqref{NCc} for the effective action.

The first, second and completely symmetric third order coefficients in the expansion \eqref{Myersexpansion} of the effective action are fixed in terms of the derivatives of $c$ by consistency conditions \cite{Ferrari:2013pi}. To get further information, we thus need to compute the completely antisymmetric third order coefficient or equivalently the three-form $F^{(3)}$ defined in \eqref{F3def}. From \eqref{s3}, we see that the determinant in \eqref{SeffboseNC} cannot contribute to the completely antisymmetric coefficient. A priori, we thus simply need
to plug the solution of \eqref{EOMDNC} to the third order in $\epsilon$ into \eqref{calSNC}. However, the algebra to do this calculation explicitly is daunting. Very fortunately, the discussion can be greatly simplified by using the following argument.

The basic idea is to note that the $D$-dependent piece \eqref{calSNC} of the effective action and thus the saddle-point equation \eqref{EOMDNC} as well depend only on the combinations $\smash{\vec Y^{2}}$ and $[X_{\mu},X_{\nu}]^{+}=\ls^{4}[\epsilon_{\mu},\epsilon_{\nu}]^{+}$ of the matrices $Y_{A}$s and $X_{\mu}$s. The same must be true after plugging $D_{\mu\nu}=\langle D_{\mu\nu}\rangle$ into $\mathcal S$. If we define
\be\label{epsrdef} \vec Y^{2} = r^{2} + \ls^{2}\epsilon_{r} = r^{2} + 2\ls^{2}\vec y\cdot\vec\epsilon + \ls^{4}\vec{\epsilon\,}^{2}\, ,\quad
\ee
the expansion of $\mathcal S$ in powers of $\epsilon$ is then most conveniently written in terms of $[\epsilon_{\mu},\epsilon_{\nu}]^{+}$
and $\epsilon_{r}$. It will actually be useful to replace $[\epsilon_{\mu},\epsilon_{\nu}]^{+}$ by a completely general self-dual matrix $M_{\mu\nu}^{+}$ in \eqref{calSNC} and \eqref{EOMDNC}, which is not necessarily a commutator, and solve the equations in term of this more general matrix. We simply have to keep in mind that $M_{\mu\nu}^{+}$ will be identified with $[\epsilon_{\mu},\epsilon_{\nu}]^{+}$ at the end of the calculation and is thus of order $\epsilon^{2}$.
The most general single-trace expansion up to order three then reads
\begin{multline}\label{Scalexpand} \mathcal S
\bigl(M_{\mu\nu}^{+},r^{2}+\ls^{2}\epsilon_{r}\bigr) = K s(r^{2}) + \ls^{2}s'(r^{2})\tr\epsilon_{r}
+\frac{\ls^{4}}{2}s''(r^{2})\tr\epsilon_{r}^{2} + \frac{\ls^{6}}{6}s'''(r^{2})\tr
\epsilon_{r}^{3}\\ + \ls^{4}s_{\mu\nu}(r^{2})\tr M_{\mu\nu}^{+}
 + \ls^{6}s'_{\mu\nu}(r^{2})\tr\epsilon_{r}
M_{\mu\nu}^{+} + O\bigl(\epsilon^{4}\bigr)\, ,
\end{multline}
where the primes denote the derivatives with respect to $r^{2}$.
The zeroth order coefficient $s(r^{2})$ is determined by the zeroth order solution \eqref{solD0NC} or equivalently \eqref{NCc},
\begin{align}\label{s0sola} s(r^{2}) &= c - i\vartheta-\frac{8\pi^{2}N}{\lambda} + N\ln r^{4} \\\label{s0solb} &= N
 \Bigl(\sqrt{1+4 \theta_{+}^{2}r^4/R^{8}}-1\Bigr)+N\ln\biggl(\frac{\sqrt{1+4 \theta_{+}^{2}r^4/R^{8}}-1}{2\theta_{+}^{2}/R^{8}}\biggr)\, .
\end{align}
Since $\mathcal S$ does not depend on $r^{2}$ and $\epsilon_{r}$ independently but only through the combination $r^{2}+\ls^{2}\epsilon_{r}$, the expansion \eqref{Scalexpand} must be invariant under the simultaneous shifts \cite{Ferrari:2013pi}
\be\label{shift1} r^{2}\rightarrow r^{2} + \ls^{2} a\, ,\quad \epsilon_{r}\rightarrow\epsilon_{r} - a\1\, ,\ee
for any real number $a$. This fixes 
the terms in $\tr\epsilon_{r}$, $\tr\epsilon_{r}^{2}$ and $\tr\epsilon_{r}^{3}$ in terms of the derivatives of $s$ and the term in $\tr\epsilon_{r}M_{\mu\nu}^{+}$ in terms of $s'_{\mu\nu}$ as indicated. To fix $s_{\mu\nu}(r^{2})$, we can then use another shift symmetry, under
\be\label{shift2} M_{\mu\nu}^{+}\rightarrow 
M_{\mu\nu}^{+} + i\xi_{\mu\nu}^{+}\, ,\quad
\theta_{\mu\nu}^{+}\rightarrow\theta_{\mu\nu}^{+}-\ls^{4}\xi_{\mu\nu}^{+}\, ,\ee
for any self-dual $\xi_{\mu\nu}^{+}$. This symmetry comes
from the fact that only the combination $\ls^{4} M_{\mu\nu}^{+}+ i\theta_{\mu\nu}^{+}$ enters in the generalized versions of the equations \eqref{calSNC} and \eqref{EOMDNC}, in which $[X_{\mu},X_{\nu}]^{+}$ has been replaced by $\ls^{4}M_{\mu\nu}^{+}$. This replacement is useful precisely because it allows to consider the symmetry \eqref{shift2}, by waiving the tracelessness condition that any commutator must satisfy. The invariance of \eqref{Scalexpand} under \eqref{shift2} then yields
\be\label{smunusol} s_{\mu\nu} = -i\frac{\partial s}{\partial\theta_{\mu\nu}^{+}} = \frac{iN}{\theta_{+}^{2}}\biggl[1-\sqrt{1+4 \theta_{+}^{2}r^4/R^{8}} \biggr]\theta_{\mu\nu}^{+}\, .
\ee
Plugging this result in \eqref{Scalexpand} for $M_{\mu\nu}^{+}=[\epsilon_{\mu},\epsilon_{\nu}]^{+}$ and using \eqref{epsrdef} immediately yields the piece
\be\label{piece} 2\ls^{6}s'_{\mu\nu}y_{A}\tr \epsilon_{A}[\epsilon_{\mu},\epsilon_{\nu}]\ee
of the effective action contributing to the three-form $F^{(3)}$ in \eqref{F3def}, from which we obtain
\be\label{F3NCsol} F^{(3)} = 4s'_{\mu\nu}y_{A}\d x^{\mu}\wedge\d
x^{\nu}\wedge\d y^{A} = \d\bigl[2 s_{\mu\nu}\d x^{\mu}\wedge\d x^{\nu}\bigr]\, .\ee
This is equivalent to the formula \eqref{NComega} for the two-form $C^{(2)}$ defined in \eqref{dc}.

\subsubsection{The emergent geometry}

In this example, there is a non-trivial contribution \eqref{NCc} to the action at order $\epsilon^0$.
As we have extensively discussed in Sections \ref{useSec} and \ref{NCsubex}, the physical content of this formula is obtained by expanding up to quadratic order in the deformation parameter $\theta_{+}$ and comparing with \eqref{c0}. This yields
\begin{equation}
  \tau= ie^{-\phi}-C_{0}= -\frac{\vartheta}{2\pi}+\frac{4i\pi N}\lambda\left( 1+
  \frac{\theta_{+}^{2}r^{4}}{2\ls^4 R^{4}}\right)+O\bigl(\ls^{-2}\theta\bigr)^3 \, .
  \label{NCtauSUGRA}
\end{equation}
To disentangle the dilaton and the axion fields from \eqref{NCtauSUGRA}, one has to be careful because the fields do not need to be real-valued in the Euclidean. It is thus convenient to rotate the $x_{4}$ coordinate to Minkowskian time which, from \eqref{NCrelation}, implies that $\theta_{34}$ is purely imaginary. After this rotation, the dilaton $\phi$ and the axion $C_0$ are real and we can then take the real and imaginary parts of \eqref{NCtauSUGRA} to get \eqref{C0NCsol} and \eqref{phiNCsol}.

Similarly, the action at third order yields \eqref{NComega} as we have shown. The physical content of this contribution is found by expanding to linear order in $\theta_{+}$, see Sections \ref{useSec} and \ref{NCsubex}.
From \eqref{dc} and \eqref{c3}, this yields
\begin{align}
  \tau B-C_2&=\frac{4i\pi N}{\lambda}\frac{ r^4}{R^4} \frac{\theta^\sd_{\mu\nu}}{\ls^2} \d x_\mu\wedge\d x_\nu+O\bigl(\ls^{-2}\theta\bigr) \, .
  \label{BC2NC}
\end{align}
To disentangle the Neveu-Schwarz and Ramond-Ramond fields $B$ and $C_2$ from \eqref{BC2NC}, we again rotate to Minkowskian signature in which $x_4$ and $\theta_{34}$ are purely imaginary and the fields $B$ and $C_2$ are real. Taking the real and imaginary parts of \eqref{BC2NC} then yields \eqref{C2NCsol} and \eqref{BNCsol}. 

As a final remark, let us note that we have also computed the effective action to the fourth order. As mentioned in Section \ref{useSec}, only the term linear in the deformation parameter $\theta$ is physical. Consistently with the supergravity solution, this linear term is found to vanish. At quadradic order in $\theta$, we find a coefficient $c_{[MN][PQ]}$ which does not factorize as in \eqref{c4}, as expected.

\subsection{The $\beta$-deformation}
\label{BetaSec}
\subsubsection{The microscopic action}

In parallel with the case of the non-commutative theory, the $\beta$-deformation can be implemented by replacing the ordinary products $fg$ appearing in the microscopic action by a $*$-product \cite{Lunin:2005jy}. Let us denote by $Q_{i}$, $1\leq i\leq 3$, the charges associated with the $\u_{1}\times\u_{2}\times\u_{3}$ subgroup of $\text{SO}(6)$ corresponding to the rotations in the 1-2, 3-4 and 5-6 planes in $\vec y$-space respectively. The charge assignments according to the $\text{SU}(4)$ quantum numbers is indicated in the Appendix \ref{NotAppSec}, Table \ref{chargesU1}. The $*$-product is then defined by
\be\label{starbeta} f * g = e^{i\pi\epsilon_{ijk}\gamma_{i}Q_{j}^{f}Q_{k}^{g}}fg\, ,\ee
where $\epsilon_{ijk}$ is the totally antisymmetric symbol, the charges $Q_{i}^{f}$ and $Q_{i}^{g}$ act on $f$ and $g$ respectively and $\gamma_{1}$, $\gamma_{2}$ and $\gamma_{3}$ are three deformation parameters that we shall assume to be real. When $\gamma_{1}=\gamma_{2}=\gamma_{3}$, $\nn=1$ supersymmetry is preserved, but supersymmetry is completely broken otherwise. In all cases, the model is conformal in the planar limit \cite{Ananth:2007px,Ananth:2006ac}.

The only terms in \eqref{Sp1} that are affected when we use the $*$-product are the Yukawa couplings $\bar\psi[\phi,\bar\psi]$ and $\tilde\chi\phi\chi$. To compute the bosonic part of the effective action, we only need $\tilde\chi\phi\chi$. According to \eqref{starproductid}, the effect of the $*$-product on this term
is equivalent to replacing the matrices $\Sigma_{A}$ by deformed versions $\tilde\Sigma_{A}$,
\be\label{starbetaeffect} \tilde\chi^{a}*\Sigma_{Aab}\phi_{A}*\chi^{b}
=\tilde\chi^{a}\tilde\Sigma_{Aab}\phi_{A}\chi^{b}\, .\ee
The explicit formulas for the matrices $\tilde\Sigma_{A}$ are given in
\eqref{Sigmabetadef6D}.

\subsubsection{The effective action}

Integrating out $q$, $\tilde q$, $\chi$ and $\tilde\chi$ from the deformed microscopic action, we get
\begin{multline}\label{SeffboseBeta} S_{\text{eff}}(X,Y,D)=K\Bigl(\frac{8\pi^{2}N}{\lambda}
 +i\vartheta\Bigr)+
\frac{4\pi^{2}N}{\ls^{4}\lambda}\tr \Bigl\{
2i\ls^{4} D_{\mu\nu}\bigl[X_{\mu},X_{\nu}\bigr] -
\bigl[X_{\mu},Y_{A}\bigr]\bigl[X_{\mu},Y_{A}\bigr]\Bigr\}\\+ \ln\Delta_{q,\tilde q} - \ln\tilde\Delta_{\chi,\tilde\chi}\, ,
\end{multline}
where
\begin{align}\label{DeltaqqBeta} \ln\Delta_{q,\tilde q} &= N\ln \det \bigl(\vec 
Y^{2}\otimes \1_{2} + i \ls ^4 D_{\mu\nu}\otimes \sigma_{\mu\nu} \bigl)\, ,\\\label{DeltaccBeta}
\ln\tilde\Delta_{\chi,\tilde\chi} & =N
 \ln \det \bigl( \tilde\Sigma_A \otimes Y_A \bigl)\, .\end{align}
The dependence of $\Seff(X,Y,D)$ on $D_{\mu\nu}$ is exactly the same as in the undeformed model studied in \cite{Ferrari:2012nw}. The solution of the saddle-point equation \eqref{saddleCBD} is thus given by \eqref{Dorder3CB} for $\vec y_{f}=\vec 0$. In particular, when we write \eqref{XYexp}, $\langle D_{\mu\nu}\rangle$ is of order $\epsilon^{2}$ and will contribute to $\Seff$ only at order four or higher in $\epsilon$.

To leading order, \eqref{SeffboseBeta} yields
\be\label{cBeta} c = \frac{8\pi^{2}N}{\lambda} + i\vartheta + 2N\ln{\vec y\,}^{2} - N\ln\det U\, ,\ee
where the matrix $U$ is defined by
\be\label{Udef} U = y_{A}\tilde\Sigma_{A}\, .\ee
The determinant of $U$ can be computed straightforwardly in terms of the polar coordinates introduced in \eqref{polardef0},
\be\label{detU} \det U = \rho_{1}^{4} + \rho_{2}^{4} + \rho_{3}^{4} + 2\cos(2\pi\gamma_{1})\rho_{2}^{2}\rho_{3}^{2}
+ 2\cos(2\pi\gamma_{2})\rho_{1}^{2}\rho_{3}^{2}
+ 2\cos(2\pi\gamma_{3})\rho_{1}^{2}\rho_{2}^{2}\, .\ee
Plugging this result in \eqref{cBeta} and using the coordinates $r_{i}$ defined in \eqref{ridef0} yields
\begin{multline}\label{cfBeta} c = \frac{8\pi^{2}N}{\lambda} + i\vartheta
\\- N \ln\Bigl[ 1 - 4 \bigl(r_{2}^{2}r_{3}^{2}\sin^{2}(\pi\gamma_{1}) +
r_{1}^{2}r_{3}^{2}\sin^{2}(\pi\gamma_{2})
+r_{1}^{2}r_{2}^{2}\sin^{2}(\pi\gamma_{3})\bigr)\Bigr]\, .
\end{multline}
Let us note that this result was also obtained in the context of standard instanton calculus in \cite{Georgiou:2006np,Durnford:2006nb}.

The effective action at first and second order is fixed in terms of the derivatives of $c$. New information is found in the completely antisymmetric coefficient at order three, which yields the three-form $F^{(3)}$ defined in \eqref{F3def}. Expanding in $\epsilon$ using \eqref{detexp}, we see that
both determinants \eqref{DeltaqqBeta} and \eqref{DeltaccBeta} contribute to the third order action, but only \eqref{DeltaccBeta} yields a completely antisymmetric term. Explicitly, we get a nice and compact result,
\be\label{F3Beta} F^{(3)} = -\frac{N}{3}\tr \bigl( 
U^{-1}\d U\wedge U^{-1}\d U\wedge U^{-1}\d U\bigr)\, .\ee
In particular, this formula makes manifest the fact that $\d F^{(3)}=0$. However, 
the evaluation of the trace on the right-hand side is extremely tedious to perform manually, because the explicit expressions for the matrix $U$ and its inverse $U^{-1}$ are very complicated. We have thus implemented the calculation in Mathematica. The resulting formulas greatly simplify when using the coordinates defined in \eqref{polardef0}, \eqref{ridef0} and \eqref{thetaphidef0}. To linear order in the deformation parameters, which is all we need to compare with supergravity, we find, for the two-form potential defined in \eqref{dc},  
\begin{multline}\label{C2linearorder}
C^{(2)} = 8 N\ls^{2}\Bigl[ \omega_{1}\wedge\bigl(\gamma_{1}\d\theta_{1} + \gamma_{2}\d\theta_{2}+\gamma_{3}\d\theta_{3}\bigr) \\- \frac{i}{4}
\bigl( \gamma_{1}r_{2}^{2}r_{3}^{2}\,\d\theta_{2}\wedge\d\theta_{3}
+ \gamma_{2}r_{3}^{2}r_{1}^{2}\,\d\theta_{3}\wedge\d\theta_{1} +
\gamma_{3}r_{1}^{2}r_{2}^{2}\,\d\theta_{1}\wedge\d\theta_{2}\bigr)\Bigr] + 
O\bigl(\gamma^{2}\bigr)\, ,
\end{multline}
where the one-form $\omega_{1}$ is defined by the condition
\be\label{domega1} \d \omega_{1} = r_{1}r_{2}r_{3}\sin\theta\,\d\theta\wedge\d\phi\, .\ee
The exact result in the supersymmetry preserving case $\gamma_{1}=\gamma_{2}=\gamma_{3}=\gamma$ is given in \eqref{omegaBetasol}. Similar formulas can be obtained in other special cases, but they are not particularly illuminating. The general formula for arbitrary finite $\gamma_{i}$s is very complicated and we shall refrain from writing it down explicitly.

\subsubsection{The emergent geometry}

Expanding \eqref{cfBeta} to quadratic order in the deformation parameters and using \eqref{c0} yields
\be\label{phiBeta} e^{-\phi} = \frac{4\pi N}{\lambda}\Bigl( 1 + \frac{1}{2}\lambda\bigl( \gamma_{1}r_{2}^{2}r_{3}^{2} + \gamma_{2}r_{3}^{2}r_{1}^{2} + \gamma_{3}r_{1}^{2}r_{2}^{2}\bigr) + O\bigl(\lambda\gamma^{4}\bigr)\Bigr)\, .\ee
When the background is a small deformation of the undeformed $\AdSS$ solution, i.e.\ when $\lambda\gamma_{i}^{2}\ll 1$,
this is a perfect match with the supergravity solution \eqref{tbeta} and \eqref{Gdef}, consistently with the discussion in Section \ref{useSec}. Similarly, \eqref{C2linearorder} and \eqref{C2C4NSRR} yield
\begin{align}
\label{BgenBeta}
\begin{split}
 B & =\frac{\lambda}{4\pi N} \im C^{(2)}\\ &= -\frac{\ls^{2}\lambda}{2\pi}\bigl(
\gamma_{1}r_{2}^{2}r_{3}^{2}\,\d\theta_{2}\wedge\d\theta_{3}
+ \gamma_{2}r_{3}^{2}r_{1}^{2}\,\d\theta_{3}\wedge\d\theta_{1} +
\gamma_{3}r_{1}^{2}r_{2}^{2}\,\d\theta_{1}\wedge\d\theta_{2}\bigr) + O\bigl(\gamma^{2}\bigr)\, ,\end{split}\\
\label{C2genBeta}
\begin{split}
C_{2} & = -\re C^{(2)}-\frac{\vartheta}{2\pi} B\\& = 
-8 N\ls^{2} \omega_{1}\wedge\bigl(\gamma_{1}\d\theta_{1} + \gamma_{2}\d\theta_{2}+\gamma_{3}\d\theta_{3}\bigr)-\frac{\vartheta}{2\pi} B + 
O\bigl(\gamma^{2}\bigr)\, .\end{split}
\end{align}
After making the $\text{SL}(2,\mathbb R)$ transformation $C_{0}\rightarrow C_{0}+\frac{\vartheta}{2\pi}$, $C_{2}\rightarrow C_{2}- \frac{\vartheta}{2\pi} B$ to generalize the solution to an arbitrary bare $\vartheta$ angle, we find again a beautiful match with the supergravity background \eqref{Bbeta} and \eqref{C2beta} in the appropriate limit.

Actually, in the present case, it seems that the discussion of Section \ref{useSec} can be slightly refined. Indeed, because the imaginary part of $c$ given by \eqref{cBeta} is a constant, it turns out that the general matrix coordinates redefinitions do not act on $\re F^{(3)}$ \cite{Ferrari:2013pi}. This three-form is thus unambiguously fixed by our microscopic calculations, even when the perturbation with respect to the undeformed conformal $\nn=4$ gauge theory is large. As a consequence, to compare with supergravity, we do not have to impose $\lambda\gamma_{i}^{2}$ to be small. The only relevant constraint is of course the validity of the supergravity solution itself, which is the weaker condition $\lambda\gamma_{i}^{4}\ll 1$ together with $\lambda\gg 1$. In this limit, we are allowed to expand the microscopic results as in \eqref{C2linearorder},
since $\gamma_{i}\ll 1$. However, we are not allowed to simplify the function $G$ defined by \eqref{Gdef} in the supergravity solution, because $\lambda\gamma_{i}^{2}$ may be large. Remarquably, we do find agreement with the microscopic prediction, because the real part of $C^{(2)}$ is related to the right-hand side of \eqref{C2beta} which does not depend on $G$!

\section{Conclusion}
\label{ConclusionSec}

We have successfully applied the framework of \cite{Ferrari:2012nw} to three non-trivial deformations of the $\nn=4$ supersymmetric gauge theory. In spite of the limitations, explained in Section \ref{useSec}, associated with the use of D-instantons to probe the geometry, we have been able to reproduce highly non-trivial features of the supergravity duals from a purely microscopic calculation. For example, equations \eqref{C2NCsol}, \eqref{BNCsol}, \eqref{BgenBeta} and \eqref{C2genBeta} reproduce intricate solutions for the form fields in supergravity. To our knowledge, this kind of information on the gravitational duals has been totally out of reach of previous field theoretic studies. 

A very large class of models, including theories for which the supergravity dual is not yet known explicitly, can a priori be studied along the same lines. In particular, the method is not limited to D-instantons. An important next step will be to study genuine quantum mechanical pre-geometric models, corresponding to the microscopic description of higher-dimensional probe branes, from which space and a non-trivial background emerge. We hope to report on the examples with D-particles and D-strings very soon \cite{emergentD0,emergentD1}.

\subsection*{Acknowledgements}

This work is supported in part by the Belgian Fonds de la Recherche
Fondamentale Collective (grant 2.4655.07) and the Belgian Institut
Interuniversitaire des Sciences Nucl\'eaires (grant 4.4511.06 and 4.4514.08).
M.M. and A.R. are Research Fellows of the Belgian Fonds de la Recherche Scientifique-FNRS.
\appendix

\section{\label{NotAppSec} Notations and conventions}
We work in Euclidean signature throughout this paper and do not distinguish upper and lower vector indices.

\subsection{Indices and transformation laws}
See table~\ref{indices}.

\begin{table}
\be\nonumber
\begin{matrix}
& \text{Spin}(4)  & 
 \text{SU}(4) & \text{U}(N) & \text{U}(K)\\
\hline 
\alpha, \beta, ... \ \text{(upper or lower)}& (1/2,0) & \mathbf 1 & \mathbf 1&\mathbf 1 \\
\dot\alpha, \dot\beta, ... \ \text{(upper or lower)}& (0,1/2) & \mathbf 1 & \mathbf 1&\mathbf 1 \\
\mu, \nu, ... & (1/2,1/2) & \mathbf 1 & \mathbf 1&\mathbf 1 \\
a, b, ... \ \text{(lower)}& (0,0) & \mathbf 4 & \mathbf 1&\mathbf 1 \\
a, b, ... \ \text{(upper)}& (0,0) & \mathbf{\bar 4} & \mathbf 1&\mathbf 1 \\
A, B, ... & (0,0) & \mathbf{6} & \mathbf 1&\mathbf 1 \\
f, f', ... \ \text{(lower)}& (0,0) & \mathbf 1 & \mathbf N&\mathbf 1 \\
f, f', ... \ \text{(upper)}& (0,0) & \mathbf 1 & \mathbf{\bar N}&\mathbf 1 \\
i, j, ... \ \text{(lower)}& (0,0) & \mathbf 1 & \mathbf 1&\mathbf K \\
i, j, ... \ \text{(upper)}& (0,0) & \mathbf 1 & \mathbf 1&\mathbf{\bar K}\\
X_{\mu i}^{\ \ j}=\ls^{2}A_{\mu i}^{\ \ j} & (1/2,1/2) & \mathbf 1 & \mathbf 1 & \textbf{Adj}\\
Y_{A i}^{\ \ j}=\ls^{2}\phi_{A i}^{\ \ j} & (0,0) & \mathbf 6 & \mathbf 1& \textbf{Adj} \\
\psi_{\alpha a i}^{\ \ \ j} =\ls^{2}\Lambda_{\alpha a i}^{\ \ \ j} & (1/2,0) & \mathbf 4 & \mathbf 1  & \textbf{Adj}\\
\bar\psi^{\dot\alpha a j}_{\ \ i} =\ls^{2}\bar\Lambda^{\dot\alpha a j}_{\ \ i} & (0,1/2) & \mathbf{\bar 4} & \mathbf 1
& \textbf{Adj} \\
D_{\mu\nu i}^{\ \ \ j} & (1,0) & \mathbf 1 & \mathbf 1 & \textbf{Adj}\\
q_{\alpha f i} & (1/2,0) & \mathbf 1 & \mathbf N  & \textbf{K}\\
\tilde q^{\alpha f i} & (1/2,0) & \mathbf 1 & \mathbf{\bar N}  & 
\mathbf{\bar K}\\
\chi^{a}_{\ f i} & (0,0) & \mathbf{\bar 4} & \mathbf{N}  & 
\mathbf{K}\\
\tilde\chi^{afi} & (0,0) & \mathbf{\bar 4} & \mathbf{\bar N}  & 
\mathbf{\bar K}
\end{matrix}
\ee
\caption{\label{indices}Conventions for the transformation laws of indices and moduli. For maximum clarity, we have indicated all the indices associated to each modulus, whereas in the main text the gauge $\text{U}(N)$ and $\text{U}(K)$ indices are usually suppressed. The representations of $\text{Spin}(4)=\text{SU}(2)_{+}\times
\text{SU}(2)_{-}$ are indicated according to the spin in each $\text{SU}(2)$ factor. The $(1/2,1/2)$ of $\text{SU}(2)_{+}\times
\text{SU}(2)_{-}$ and the $\mathbf 6$ of $\text{SU}(4)=\text{Spin}(6)$ correspond to the fundamental representations of $\text{SO}(4)$ and $\text{SO}(6)$ respectively.}   
\end{table}

\subsection{Four-dimensional algebra}
With the standard Pauli matrices
\begin{equation}
  \sigma_1=\begin{pmatrix}0 & 1\\1 & 0\end{pmatrix},\quad \sigma_2=\begin{pmatrix}0 & -i\\i & 0\end{pmatrix}, \quad \sigma_3=\begin{pmatrix}1 & 0\\0 & -1\end{pmatrix} \, ,
  \label{paulidef}
\end{equation}
we can define
\begin{equation}
  \sigma_{\mu\alpha\dot\alpha}=(\vec\sigma,-i\1_{2})_{\alpha\dot\alpha}\,, \quad \bar\sigma_\mu^{\dot\alpha\alpha}=(-\vec\sigma,-i\1_{2})^{\dot\alpha\alpha}
  \label{sigma1def}
\end{equation}
and
\begin{equation}
  \sigma_{\mu\nu}=\frac14(\sigma_{\mu}\bar\sigma_{\nu}-\sigma_{\nu}\bar\sigma_{\mu})\,, \quad 
  \bar\sigma_{\mu\nu}=\frac14(\bar\sigma_{\mu}\sigma_{\nu}-\bar\sigma_{\nu}\sigma_{\mu})\, .
  \label{sigma2def}
\end{equation}
The following identity is very useful:
\begin{equation}
  \sigma_{\mu\nu}\sigma_{\rho\kappa}=\frac14(-\epsilon_{\mu\nu\rho\kappa}+\delta_{\nu\rho}\delta_{\mu\kappa}-\delta_{\mu\rho}\delta_{\nu\kappa})\1_{2} + \left( \delta_{\kappa [\nu}\sigma_{\mu]\rho}-\delta_{\rho[\nu} \sigma_{\mu]\kappa} \right) \, ,
  \label{2sigmaid}
\end{equation}
where $\epsilon_{\mu\nu\rho\sigma}$ is the completely antisymmetric tensor with $\epsilon_{1234}=+1$.

We denote by an upper ``+'' sign the projection of an antisymmetric tensor on its self-dual part,
\begin{equation}
  a^+_{\mu\nu}=\frac12(a_{\mu\nu}+\frac12\epsilon_{\mu\nu\rho\kappa}a_{\rho\kappa}) \, .
  \label{sddef}
\end{equation}
With these definitions $\sigma_{\mu\nu}$ is self-dual,
\begin{align}
  \sigma_{\mu\nu}=\sigma_{\mu\nu}^+ \, .
  \label{sigmasd}
\end{align}
Let us finally mention the following useful identities,
\begin{align}\det(\1_2+a_{\mu\nu}\sigma_{\mu\nu})& =1+a_+^2 \, ,
  \label{detsigmaid}\\
\bigl( \1_{2}+a_{\mu\nu}\sigma_{\mu\nu} \bigr)^{-1}&=
\frac{\1_{2}-a_{\rho\sigma}\sigma_{\rho\sigma}}{1+a_+^2}\, \cvp
  \label{inv1sigma}
\end{align}
where
\be\label{defaplus}
a_+^2 = a^+_{\mu\nu}a^+_{\mu\nu}\, .
\ee
\subsection{Six-dimensional algebra}
\subsubsection{Undeformed case}

We define
\begin{align}\nonumber
&\Sigma_{1}=
\begin{pmatrix} 0&-1&0&0\\1&0&0&0\\0&0&0&1\\0&0&-1&0\end{pmatrix}\, ,\
\Sigma_{2}=
\begin{pmatrix} 0&-i&0&0\\i&0&0&0\\0&0&0&-i\\0&0&i&0\end{pmatrix}\, ,\
\Sigma_{3}=
\begin{pmatrix} 0&0&-1&0\\0&0&0&-1\\1&0&0&0\\0&1&0&0\end{pmatrix}\, ,\\
\label{Sigmadef6D}
&\Sigma_{4}=
\begin{pmatrix} 0&0&-i&0\\0&0&0&i\\i&0&0&0\\0&-i&0&0\end{pmatrix}\, ,\
\Sigma_{5}=
\begin{pmatrix} 0&0&0&-1\\0&0&1&0\\0&-1&0&0\\1&0&0&0\end{pmatrix}\, ,\
\Sigma_{6}=
\begin{pmatrix} 0&0&0&-i\\0&0&-i&0\\0&i&0&0\\i&0&0&0\end{pmatrix}
\end{align}
and
\be\label{Sigmabardef6D} \bar\Sigma_{A}= \Sigma_{A}^{\dagger}\, .\ee
These matrices satisfy the algebra
\be
\Sigma_A\bar\Sigma_B+\Sigma_B\bar\Sigma_A =2\delta_{AB}\1_4 
\ee
as well as the relations
\be\label{relSigma} \bar\Sigma_{A}^{ab}= \frac{1}{2}\epsilon^{abcd}
\Sigma_{Acd}\, ,\quad 
\Sigma_{Aab}= \frac{1}{2}\epsilon_{abcd}\bar\Sigma_{A}^{cd}\ee
where the $\epsilon$s are completely antisymmetric symbols with $\epsilon_{1234}=\epsilon^{1234}=+1$.
Euclidean six-dimensional Dirac matrices, satisfying
\be\label{Cliff6D} \bigl\{\Gamma_{A},\Gamma_{B}\bigr\} = 2\delta_{AB}\, ,\ee
can then be defined by
\be\label{gamma6Ddef} \Gamma_{A} = 
\begin{pmatrix}
0 & \Sigma_{A}\\ \bar\Sigma_{A} & 0
\end{pmatrix}\, .
\ee

If $\vec v = (v_{A})_{1\leq A\leq 6}$ is a six-dimensional vector, one can check that
\begin{align}
\label{detsigmaa}\det(v_{A}\Sigma_A)&={\vec v\,}^4\,,\\
(v_{A}\Sigma_A)^{-1}&=\frac{v_{A}\bar \Sigma_A}{{\vec v\,}^{2}}\, \cdotp
\end{align}
In Sections \ref{CBSec} and \ref{NCSec} of the main text, we have to compute the expansion of some determinants of the form
\be\label{appdetex} \ln\det\bigl(\Sigma_{A}\otimes (v_{A} + \ls^{2}\epsilon_{A})\bigr) = \ln {\vec v\,}^4 + \sum_{k=1}^\infty\frac{(-1)^k}{k}\tr\left( \left(v_A \Sigma_A\right)^{-1}\Sigma_B \otimes \epsilon_{B} \right)^k = \sum_{k=0}^{\infty}t^{(k)} \, .\ee
Up to order five, this is done by using the trace formulas in \cite{Ferrari:2012nw}, which yield
\begin{align}
  \label{s1}t^{(1)} &= -\frac{4}{ v^{2}}\tr_\UK (\vec v\cdot\vec\epsilon\,)\,,\\
  \label{s2}t^{(2)} &= \frac2{{\vec v \,}^{4}}\tr_\UK \left[ 2(\vec v\cdot\vec\epsilon\,)^2 - {\vec v \,}^2 {\vec\epsilon\,}^2 \right] \,,\\
  \label{s3}t^{(3)} &= -\frac4{3 {\vec v \,}^{6}}\tr_\UK \left[ 4(\vec v\cdot\vec\epsilon\,)^3 - 3 {\vec v \,}^2 (\vec v\cdot\vec\epsilon\,){\vec\epsilon\,}^2 \right] \,,\\
  \label{s4}t^{(4)} &= \frac{8}{ {\vec v \,}^{8}} \tr_\UK \left[ (\vec v\cdot\vec\epsilon\,)^4 - {\vec v \,}^2 (\vec v\cdot\vec\epsilon\,)^2 {\vec\epsilon\,}^2 + \frac 14 {\vec v \,}^4 {\vec\epsilon\,}^4 - \frac18 {\vec v \,}^4 \epsilon_A\epsilon_B\epsilon_A\epsilon_B \right] \,,\\
  \label{s5}t^{(5)} &= -\frac{4}{{\vec v \,}^{10}} \tr_\UK \biggl[ \frac{16}5 (\vec v\cdot\vec\epsilon\,)^5 - 4{\vec v \,}^2 (\vec v\cdot\vec\epsilon\,)^3 {\vec\epsilon\,}^2 +
  \notag\\& {\vec v \,}^4 \left( \vec v\cdot\vec \epsilon\, {\vec\epsilon\,}^4 - \vec v\cdot\vec \epsilon \, \epsilon_B\epsilon_C\epsilon_B\epsilon_C + \vec v\cdot\vec \epsilon \, \epsilon_B {\vec\epsilon\,}^2 \epsilon_B \right) +
\frac i5 {\vec v \,}^4 v_A \epsilon_{A_1\cdots A_5A}\epsilon_{A_1}\cdots \epsilon_{A_5} \biggr] \,.
\end{align}

Weyl spinors $\lambda_{a}$ and $\psi^{a}$ in the $\mathbf 4$ and $\mathbf{\bar 4}$ representations of the
rotation group $\text{Spin}(6)=\text{SU}(4)$ transform under a 
six-dimensional rotation parametrized by the antisymmetric matrix $\Omega$, $\delta x_{A}= -\Omega_{AB}x_{B}$, as
\be\label{spinors6Dtl} \delta\lambda_{a} = -\frac{1}{2}\Omega_{AB}
\Sigma_{ABa}^{\ \ \ \ b}\lambda_{b}\, ,\quad
\delta\psi^{a} = -\frac{1}{2}\Omega_{AB}\bar\Sigma_{AB\ b}^{\ \ \ a}
\psi^{b}\, ,\ee
where the generators of the rotation group are defined by
\be\label{gen6Ddef} \Sigma_{AB}=\frac{1}{4}\bigl( \Sigma_{A}\bar
\Sigma_{B} - \Sigma_{B}\bar\Sigma_{A}\bigr)\, , \quad
\bar\Sigma_{AB}=\frac{1}{4}\bigl( \bar\Sigma_{A}
\Sigma_{B} - \bar\Sigma_{B}\Sigma_{A}\bigr)\, .\ee
This yields in particular the charges under the $\u_{1}\times\u_{2}\times\u_{3}$ subgroup of $\text{SO}(6)$ corresponding to rotations in the 1-2, 3-4 and 5-6 planes respectively, see Table \ref{chargesU1}.

\subsubsection{$\beta$-deformed case}

The $\u_{i}$ charges in Table \ref{chargesU1} are used to compute the $*$-product in Section \ref{BetaSec}. In particular, deformed $\Sigma_{A}$ matrices can be defined by the identity
\be\label{starproductid} \psi_{1}^{a}*\phi_{A}*\psi_{2}^{b}\,\Sigma_{Aab} = 
\psi_{1}^{a}\phi_{A}\psi_{2}^{b}\,\tilde\Sigma_{Aab}\, .\ee
Explicitly, we have
\begin{align}\nonumber
\tilde\Sigma_{1}&=
\begin{pmatrix} 0&-i^{\gamma_{1}-\gamma_{2}}&0&0\\i^{-\gamma_{1}+
\gamma_{2}}&0&0&0\\0&0&0& i^{-\gamma_{1}-\gamma_{2}} \\0&0&
-i^{\gamma_{1}+\gamma_{2}}&0\end{pmatrix}\, ,\\\nonumber
\tilde\Sigma_{2} &=
\begin{pmatrix} 0&i^{\gamma_{1}-\gamma_{2}-1}&0&0\\i^{-\gamma_{1}+\gamma_{2}+1}&0&0&0\\0&0&0&i^{-\gamma_{1}-\gamma_{2}-1}\\0&0&
i^{\gamma_{1}+\gamma_{2}+1}&0\end{pmatrix}\, ,\\\nonumber
\tilde\Sigma_{3}&=
\begin{pmatrix} 0&0&-i^{-\gamma_{1}+\gamma_{3}}&0\\0&0&0&
-i^{\gamma_{1}+\gamma_{3}}\\i^{\gamma_{1}-\gamma_{3}}&0&0&0\\0&i^{-\gamma_{1}-\gamma_{3}}&0&0\end{pmatrix}\, ,\\
\label{Sigmabetadef6D}
\tilde\Sigma_{4} & =
\begin{pmatrix} 0&0&i^{-\gamma_{1}+\gamma_{3}-1}&0\\0&0&0&i^{\gamma_{1}+\gamma_{3}+1}\\i^{\gamma_{1}-\gamma_{3}+1}&0&0&0\\0&i^{-\gamma_{1}-\gamma_{3}-1}&0&0\end{pmatrix}\, ,\\\nonumber
\tilde\Sigma_{5}& =
\begin{pmatrix} 0&0&0&-i^{\gamma_{2}-\gamma_{3}}\\0&0&i^{-\gamma_{2}-\gamma_{3}}&0\\0&-i^{\gamma_{2}+\gamma_{3}}&0&0\\i^{-\gamma_{2}+\gamma_{3}}&0&0&0\end{pmatrix}\, ,\\\nonumber
\tilde\Sigma_{6} &=
\begin{pmatrix} 0&0&0&i^{\gamma_{2}-\gamma_{3}-1}\\0&0&i^{-\gamma_{2}-\gamma_{3}-1}&0\\0&i^{\gamma_{2}+\gamma_{3}+1}&0&0\\i^{-\gamma_{2}+\gamma_{3}+1}&0&0&0\end{pmatrix}\, .
\end{align}
\begin{table}
\be\nonumber
\begin{array}{c|cccrrrrrrrr}
\ & y_{1}+ i y_{2} & y_{3}+ i y_{4} & y_{5}+ i y_{6} & \lambda_{1} & \lambda_{2} & \lambda_{3} & \lambda_{4} & \psi^{1} & \psi^{2} & \psi^{3}
& \psi^{4}\\\hline\\[-12pt]
\u_{1}\ & 1 & 0 & 0 & \frac{1}{2} & \frac{1}{2} & -\frac{1}{2} &-\frac{1}{2}  & -\frac{1}{2} & -\frac{1}{2} & \frac{1}{2} & \frac{1}{2} 
\\[3pt]
 \u_{2}\ & 0 & 1 & 0 & \frac{1}{2} & -\frac{1}{2} & \frac{1}{2} &-\frac{1}{2}  &-\frac{1}{2}  &\frac{1}{2}  &-\frac{1}{2}  & \frac{1}{2} 
 \\[3pt]
 \u_{3}\ & 0 & 0 & 1 &\frac{1}{2}  &-\frac{1}{2}  &-\frac{1}{2}  &\frac{1}{2}  &-\frac{1}{2}  &\frac{1}{2}  &\frac{1}{2}  &-\frac{1}{2}  
\end{array}
\ee\caption{\label{chargesU1}Charges under $\u_{1}\times\u_{2}\times\u_{3}\subset\text{SO}(6)$. The spinors $\lambda_{a}$ and $\psi^{a}$ are arbitrary spinors in the $\mathbf 4$ and $\mathbf{\bar 4}$ representations of $\text{Spin}(6)$ respectively.}
\end{table}
\section{Myers' non-abelian D-instanton action}
\label{appB}

Myers' non-abelian D-instanton action \cite{Myers:1999ps}, in the expansion \eqref{Myersexpansion} up to order five, is given in terms of the type IIB supergravity fields by the following formulas \cite{Ferrari:2012nw},
\begin{align}\nonumber
S_{\text{eff}}^{(0)} & = -2i\pi K\tau\, ,\\\nonumber
S_{\text{eff}}^{(1)} & = -2i\pi \ls^{2}\partial_{M}\tau\tr \epsilon_{M}\, ,\\
\nonumber
S_{\text{eff}}^{(2)} & = -i\pi\ls^{4}\partial_{M}\partial_{N}\tau\tr
\epsilon_{M}\epsilon_{N}\, ,\\\nonumber
S_{\text{eff}}^{(3)} & =\bigl(-\frac{i\pi}{3}\ls^{6}\partial_{M}\partial_{N}\partial_{P}\tau - 
2\pi\ls^{4}\partial_{[M}(\tau B - C_{2})_{NP]}\bigr)\tr\epsilon_{M}\epsilon_{N}\epsilon_{P} \, ,\\\label{Mexpand}
S_{\text{eff}}^{(4)} & =\bigl( -\frac{i\pi}{12}\ls^{8}
\partial_{M}\partial_{N}\partial_{P}\partial_{Q}\tau
- \frac{3\pi}{2}\ls^{6}\partial_{M}
\partial_{[N}(\tau B - C_{2})_{PQ]}\\\nonumber
& \hskip 4cm-\pi\ls^{4}e^{-\Phi}(G_{MP}G_{NQ} - G_{MQ}G_{NP})\bigr)
\tr\epsilon_{M}\epsilon_{N}\epsilon_{P}\epsilon_{Q}\, ,
\\\nonumber
S_{\text{eff}}^{(5)} & = \Bigl(-\frac{i\pi}{60}\ls^{10}\partial_{M}\partial_{N}\partial_{P}\partial_{Q}\partial_{R}\tau 
- \frac{\pi}{3}\ls^{8}\partial_{P}\partial_{Q}\partial_{R}(\tau B - C_{2})_{MN}\\\nonumber &\hskip 4cm
- \pi\ls^{6}\partial_{R}\bigl(e^{-\Phi}
(G_{MP}G_{NQ} - G_{MQ}G_{NP})\bigr)\\\nonumber &\hskip 2cm
-i\pi\ls^{6}\partial_{[M}(C_{4}+ C_{2}\wedge B - \frac{\tau}{2}
B\wedge B)_{NPQR]}\Bigr)
\tr\epsilon_{M}\epsilon_{N}\epsilon_{P}\epsilon_{Q}\epsilon_{R}\, .
\end{align}
\section{Some type IIB supergravity backgrounds}
\label{SolSUGRA}

We review in this appendix the known supergravity backgrounds dual to the non-com\-mu\-ta\-ti\-ve and $\beta$-deformed Euclidean $\nn=4$ super Yang-Mills theories studied in the main text. We use the standard relation between the radius $R$ and the 't~Hooft coupling $\lambda$,
\be\label{Rlambdarelapp} R^{4}=\alpha'^{2}\lambda = \frac{\ls^{4}\lambda}{4\pi^{2}}\,\cdotp\ee
The backgrounds are written at zero bare $\vartheta$ angle. The solutions at non-zero $\vartheta$ can be obtained by performing the $\text{SL}(2,\mathbb R)$ transformation $C_{0}\rightarrow C_{0} + \frac{\vartheta}{2\pi}$, $C_{2}\rightarrow C_{2}-
\frac{\vartheta}{2\pi} B$ and $C_{4}\rightarrow C_{4} + \frac{\vartheta}{4\pi}B\wedge B$, which automatically yields a new solution to the supergravity equations of motion.

\subsection{\label{App2}The dual to the non-commutative gauge theory}

The gravitational dual of the non-commutative deformation of the $\nn=4$ super Yang-Mills theory was derived by Hashimoto, Itzhaki, Maldacena and Russo in \cite{Hashimoto:1999ut,Maldacena:1999mh}.\footnote{Our formulas can be matched with those in \cite{Maldacena:1999mh} by making the replacements $R^2\rightarrow\alp R^2$, $\theta_{12}\rightarrow\tilde b' /(2\pi)$, $\theta_{34}\rightarrow\tilde b/(2\pi)$, $r\rightarrow\alp R^2 u$, $\lambda/(4\pi N)\rightarrow \hat g$ and $C_0\rightarrow-\chi$, $C_2\rightarrow-A$, $F_5\rightarrow-F$.} With non-vanishing non-commutative parameters $\theta_{12}=-\theta_{21}$ and $\theta_{34}=-\theta_{43}$, the solution for the string-frame metric and the other supergravity fields reads
\begin{align}
  \d s^2&=\frac{r^2}{R^2}\left[ \frac{ \d x_1^2+\d x_2^2 }{\Delta_{12}}+\frac{ \d x_3^2+\d x_4^2 }{\Delta_{34}} \right]+\frac{R^2}{r^2}\d r^2+R^2\d\Omega_5^2 \, , \label{metricNCsugra} \\
  e^{-\phi}&=\frac{4\pi N}{\lambda}\sqrt{\Delta_{12}\Delta_{34}} \, , \label{phiNCsugra} \\
  B&=\frac{r^4}{R^4}\left( \frac{\theta_{12}}{\ls^2}\frac{\d x_1\wedge\d x_2}{\Delta_{12}}+ \frac{\theta_{34}}{\ls^2}\frac{\d x_3\wedge\d x_4}{\Delta_{34}}  \right) \, , \label{BNCsugra} \\
  C_0&=-\frac{4i\pi N}{\lambda}\frac{\theta_{12}\theta_{34}}{\ls^4}\frac{r^4}{R^4} \, \cvp \label{C0NCsugra} \\
  C_2&=-\frac{4i\pi N}{\lambda}\frac{r^4}{R^4}
  \biggl( \frac{\theta_{34}}{\ls^{2}}
    \frac{\d x_1\wedge\d x_2}{\Delta_{12}}+ 
    \frac{\theta_{12}}{\ls^{2}}
  \frac{\d x_3\wedge\d x_4}{\Delta_{34}} \biggr) \, ,
  \label{C2NCsugra} \\\label{C4NCsugra}
  C_4 &= \frac{16 \pi r^2}{R^3}\omega_4 -4i\pi\frac{r^6}{R^6}
\frac{\d x_1 \wedge \d x_2 \wedge \d x_3 \wedge \d x_4}{\Delta_{12}\Delta_{34}}\,\cvp
\end{align}
where the functions $\Delta_{12}$ and $\Delta_{34}$ are defined by
\be\label{SolNCDelta} 
\Delta_{12}=1+\left( \frac{\theta_{12}}{\ls^2} \right)^2\frac{r^4}{R^4} \, \cvp \quad \Delta_{34}=1+\biggl( \frac{\theta_{34}}{\ls^2} \biggr)^2\frac{r^4}{R^4}\, \cdotp
\ee
The $x_{1}$, $x_{2}$, $x_{3}$ and $x_{4}$ are the world-volume coordinates on which the gauge theory live, $r$ is the transverse radial coordinate, expressed in terms of the six transverse coordinates $\vec y = (y_{A})_{1\leq A\leq 6}$ as $r^2 = |\vec y|^2$, $\d\Omega_{5}^{2}$ is the metric on the five-dimensional round sphere of radius one and
$\omega_{4}$ is a four-form defined in terms of the volume form
\be\label{defomega5}
	\omega_{\Sfive} = \frac{1}{5!}\frac{R^{5}y_{F}}{r^{6}}\epsilon_{ABCDEF}\,\d y_{A}\wedge\cdots\wedge\d y_{E}
\ee
on $\Sfive$ of radius $R$ by
\be\label{defgamma5}
	\d \omega_4 = \omega_{\Sfive}\, .
\ee

The consistency of the supergravity approximation for the above solution requires as usual $\lambda\gg 1$.
In the far infrared region $r\ll R\ls/\sqrt{\theta}\sim \ls^{2}\lambda^{1/4}/\sqrt{\theta}$, the solution is a small deformation of the usual $\AdSS$ background and can be compared with the microscopic calculations presented in the main text. On the other hand, in the far ultraviolet region $r \gg R\ls/\sqrt \theta$, the metric \eqref{metricNCsugra} approximates another $\AdSS$ space, with a new radial coordinate $\tilde r=1/r$. Thus there is no conformal boundary at infinity, which signals that the non-commutative theory is not a standard UV-complete quantum field theory.

\subsection{\label{CCC}The dual to the $\beta$-deformed theory}

The gravitational dual of the $\beta$-deformed $\nn=4$ super Yang-Mills theory was derived by Lunin and Maldacena in \cite{Lunin:2005jy} in the $\nn=1$ supersymmetry preserving case $\gamma_{1}=\gamma_{2}=\gamma_{3}$ and generalized by Frolov in \cite{Frolov:2005dj} to arbitrary deformation parameters $\gamma_{1}$, $\gamma_{2}$ and $\gamma_{3}$. The solution for the string-frame metric and the other non-trivial supergravity fields reads
\begin{align}\label{metbeta}
\d s^{2} & = \frac{r^{2}}{R^{2}}\d x_{\mu}\d x_{\mu}
+\frac{R^{2}}{r^{2}}\d r^{2} + R^{2}\d\tilde\Omega_{5}^{2}\, ,\\
\label{tbeta}
e^{-\phi} &=\frac{4\pi N}{\lambda\sqrt{G}}\, \cvp\\\label{Bbeta}
B&=-\frac{\ls^{2}\lambda}{2\pi}\,   G\,
 \bigl(\gamma_{3}r_{1}^{2}r_{2}^{2}\d\theta_{1}\wedge
\d\theta_{2}+\gamma_{2}r_{1}^{2}r_{3}^{2}\d\theta_{3}\wedge
\d\theta_{1}+
\gamma_{1}r_{2}^{2}r_{3}^{2}\d\theta_{2}\wedge\d\theta_{3}\bigr)\, ,\\
\label{C2beta} C_{2}&= -8N\ls^{2}\,\omega_{1}\wedge\bigl(
\gamma_{1}\d\theta_{1}+\gamma_{2}\d\theta_{2}+\gamma_{3}\d\theta_{3}\bigr)
\, ,\\\label{C4beta}
C_{4} &= \frac{4N\ls^{4}}{\pi}\bigl(G\,\omega_{1}\wedge\d\theta_{1}\wedge\d\theta_{2}\wedge\d\theta_{3}-i\omega_{4}\bigr)\, .\end{align}
The coordinates $x_{\mu}$, $1\leq\mu\leq 4$, can be viewed as the world-volume coordinates of the background D3-branes. 
The coordinate $r$ is the usual transverse radial coordinate, expressed in terms of the six transverse coordinates $\vec y = (y_{A})_{1\leq A\leq 6}$ as $r^{2}=\vec y^{2}$. The coordinates $(r_{i},\theta_{i})_{1\leq i\leq 3}$ are defined by the relations
\begin{alignat}{3}\nonumber y_{1}&= \rho_{1}\cos\theta_{1}\, ,\quad &
y_{3} &=\rho_{2}\cos\theta_{2}\, , \quad & y_{5} &=\rho_{3}\cos\theta_{3}\, ,
\\\label{polardef}
y_{2} &=\rho_{1}\sin\theta_{1}\, ,&
y_{4} &=\rho_{2}\sin\theta_{2}\, ,& y_{6} &=\rho_{3}\sin\theta_{3}
\end{alignat}
and
\be\label{ridef} r_{i} =\frac{\rho_{i}}{\sqrt{\rho_{1}^{2}+\rho_{2}^{2}+\rho_{3}^{2}}}=\frac{\rho_{i}}{|\vec y|}\,\cvp\quad r_{1}^{2}+r_{2}^{2}+r_{3}^{2}=1\, .
\ee
The function $G$ is given by 
\be\label{Gdef} \frac{1}{G} = 1+\lambda\bigl(\gamma_{1}^{2}r_{2}^{2}r_{3}^{2}
+\gamma_{2}^{2}r_{1}^{2}r_{3}^{2}+\gamma_{3}^{2}r_{1}^{2}r_{2}^{2}\bigr)\, .\ee
The metric \eqref{metbeta} describes an $\AdS\times\tilde{\text{S}}{}^{5}$ geometry for a deformed five-sphere $\tilde{\text{S}}{}^{5}$ endowed with the metric
\be\label{defS5met} \d\tilde\Omega_{5}^{2}=\sum_{i=1}^{3}\bigl( \d r_{i}^{2} + G\, r_{i}^{2}\d\theta_{i}^{2}\bigr) + \lambda G\, r_{1}^{2}r_{2}^{2}r_{3}^{2}
\Bigl(\sum_{i=1}^{3}\gamma_{i}\d\theta_{i}\Bigr)^{2}\, .
\ee
Defining the angles $\theta$ and $\phi$ by
\be\label{thetaphidef} r_{1}=\sin\theta\cos\phi\, ,\quad r_{2}=\sin\theta\sin\phi\, ,\quad r_{3}=\cos\theta\, ,\ee
the one-form $\omega_{1}$ in \eqref{C2beta} and \eqref{C4beta} satisfies
\be\label{domeg1}\d\omega_{1} = r_{1}r_{2}r_{3}\, \sin\theta\,\d\theta\wedge\d\phi\ee
and can be chosen to be
\be\label{omeg1form} \omega_{1}=\frac{1}{4}\sin^{4}\theta\cos\phi\sin\phi\,\d\phi\, .\ee
The four-form $\omega_{4}$ in \eqref{C4beta} satisfies
\be\label{omega4cond}\d\omega_{4}= \omega_{\AdS}\, ,\ee
where 
\be \omega_{\AdS} = \frac{1}{R^{8}}\, r^{3}\d x_{1}\wedge\cdots\wedge\d x_{4}\wedge
\d r
\ee
is the volume form on the unit radius $\AdS$ space. Explicitly, one can choose
\be\label{omega4form}\omega_{4} = \frac{1}{4R^{8}}r^{4}\d x_{1}\wedge\cdots\wedge\d x_{4}\, .\ee
Changes of $\omega_{1}$ and $\omega_{4}$ by exact forms correspond to a supergravity gauge transformation.

The $\beta$-deformed theory is conformal in the planar limit, which explains the fact that the $\AdS$ factor in the metric \eqref{metricNCsugra} is undeformed. The consistency of the supergravity approximation requires, on top of the usual condition $\lambda\gg 1$, that $\gamma_{i}^{4}\lambda\ll 1$, as can be checked by evaluating the curvature of the deformed sphere \eqref{defS5met}. In particular, the $\gamma_{i}$s must be very small. This explains why the periodicity in the deformation parameters, $(\gamma_{1},\gamma_{2},\gamma_{3})\equiv (\gamma_{1}+n_{1},\gamma_{2}+n_{2},\gamma_{3}+n_{3})$ for any integers $n_{1}$, $n_{2}$, $n_{3}$, which is manifest in the microscopic theory and in particular in the effective action computed in Section \ref{BetaSec}, cannot be seen in the supergravity solution. Finally, let us note that the background is a small deformation of the usual $\AdSS$ solution when $\gamma_{i}^{2}\lambda\ll 1$, a condition often used in the main text.

\bibliographystyle{utphys}
\bibliography{emergentbiblio}
\end{document}